\documentclass[twocolumn]{aastex62}
\usepackage{amsmath}
\usepackage{units}
\usepackage{url}

\newcommand{\E}[1]{\times10^{#1}}
\newcommand{\msol}{ \, M_\sun }

\newcommand{\bi}{\begin{itemize}}
\newcommand{\ei}{\end{itemize}}
\newcommand{\commentOut}[1]{}

\shortauthors{Shen et al.}

\begin{document}

\title{\bf \Large{Three Hypervelocity White Dwarfs in \emph{Gaia} DR2: Evidence for Dynamically Driven Double-Degenerate Double-Detonation Type~Ia~Supernovae}}

\author{Ken J. Shen}
\affiliation{Department of Astronomy and Theoretical Astrophysics Center, University of California, Berkeley, CA 94720, USA}

\author{Douglas Boubert}
\affiliation{Institute of Astronomy, University of Cambridge, Madingley Rise, Cambridge, CB3 0HA, United Kingdom}

\author{Boris T. G\"{a}nsicke}
\affiliation{Department of Physics, University of Warwick, Coventry CV4 7AL, UK}

\author{Saurabh W. Jha}
\affiliation{Department of Physics and Astronomy, Rutgers, the State University of New Jersey, 136 Frelinghuysen Road, Piscataway, NJ 08854, USA}

\author{Jennifer E.\ Andrews}
\affiliation{Steward Observatory, University of Arizona, 933 N. Cherry Ave., Tucson, AZ 85721, USA}

\author{Laura Chomiuk}
\affiliation{Department of Physics and Astronomy, Michigan State University, East Lansing, MI 48824, USA}

\author{Ryan J. Foley}
\affiliation{Department of Astronomy and Astrophysics, University of California, Santa Cruz, CA 95064, USA}

\author{Morgan Fraser}
\affiliation{School of Physics, O'Brien Centre for Science North, University College Dublin, Belfield, Dublin 4, Ireland}

\author{Mariusz Gromadzki}
\affiliation{Warsaw University Astronomical Observatory, Al.\ Ujazdowskie 4, PL-00-478 Warszawa, Poland}

\author{James Guillochon}
\affiliation{Harvard-Smithsonian Center for Astrophysics, 60 Garden Street, Cambridge, MA 02138, USA}

\author{Marissa M.\ Kotze}
\affiliation{South African Astronomical Observatory, PO Box 9, Observatory, 7935, South Africa}
\affiliation{Southern African Large Telescope, PO Box 9, Observatory, 7935, South Africa}

\author{Kate Maguire}
\affiliation{Astrophysics Research Centre, School of Mathematics and Physics, Queen's University Belfast, Belfast BT7 1NN, UK}

\author{Matthew R. Siebert}
\affiliation{Department of Astronomy and Astrophysics, University of California, Santa Cruz, CA 95064, USA}

\author{Nathan Smith}
\affiliation{Steward Observatory, University of Arizona, 933 N. Cherry Ave., Tucson, AZ 85721, USA}

\author{Jay Strader}
\affiliation{Department of Physics and Astronomy, Michigan State University, East Lansing, MI 48824, USA}

\author{Carles Badenes}
\affiliation{Department of Physics and Astronomy and Pittsburgh Particle Physics, Astrophysics and Cosmology Center (PITT PACC), University of Pittsburgh, 3941 O'Hara Street, Pittsburgh, PA 15260, USA}
\affiliation{Institut de Ci\`{e}ncies del Cosmos (ICCUB), Universitat de Barcelona (IEEC-UB), Mart\'{i} Franqu\'{e}s 1, E08028 Barcelona, Spain}

\author{Wolfgang E. Kerzendorf}
\affiliation{European Southern Observatory, Karl-Schwarzschild-Strasse 2, 85748 Garching bei M\"{u}nchen, Germany}

\author{Detlev Koester}
\affiliation{Institut f\"{u}r Theoretische Physik und Astrophysik, Universit\"{a}t Kiel, 24098 Kiel, Germany}

\author{Markus Kromer}
\affiliation{Zentrum f\"{u}r Astronomie der Universit\"{a}t Heidelberg, Institut f\"{u}r Theoretische Astrophysik, Philosophenweg 12, D-69120 Heidelberg, Germany}
\affiliation{Heidelberger Institut f\"{u}r Theoretische Studien, Schloss-Wolfsbrunnenweg 35, D-69118 Heidelberg, Germany}

\author{Broxton Miles}
\affiliation{Department of Physics, North Carolina State University, Raleigh NC 27695, USA}

\author{R\"{u}diger Pakmor}
\affiliation{Heidelberger Institut f\"{u}r Theoretische Studien, Schloss-Wolfsbrunnenweg 35, D-69118 Heidelberg, Germany}

\author{Josiah Schwab}
\altaffiliation{Hubble Fellow.}
\affiliation{Department of Astronomy and Astrophysics, University of California, Santa Cruz, CA 95064, USA}

\author{Odette Toloza}
\affiliation{Department of Physics, University of Warwick, Coventry CV4 7AL, UK}

\author{Silvia Toonen}
\affiliation{Anton Pannekoek Institute for Astronomy, University of Amsterdam, 1090 GE Amsterdam, The Netherlands}

\author{Dean M.\ Townsley}
\affiliation{Department of Physics and Astronomy, University of Alabama, Tuscaloosa, AL, USA}

\author{Brian J.\ Williams}
\affiliation{Space Telescope Science Institute, Baltimore, MD 21218, USA}

\correspondingauthor{Ken J. Shen}
\email{kenshen@astro.berkeley.edu}

\begin{abstract}

Double detonations in double white dwarf (WD) binaries undergoing unstable mass transfer have emerged in recent years as one of the most promising Type Ia supernova (SN~Ia) progenitor scenarios.  One potential outcome of this ``dynamically driven double-degenerate double-detonation'' (D$^6$) scenario is that the companion WD survives the explosion and is flung away with a  velocity equal to its  $>\unit[1000]{km \, s^{-1}}$ pre-SN orbital velocity.  We perform a search for these hypervelocity runaway WDs using \emph{Gaia}'s second data release.  In this paper, we discuss seven candidates followed up with ground-based instruments.  Three sources are likely to be some of the fastest known stars in the Milky Way, with total Galactocentric velocities between $1000$ and $\unit[3000]{km \, s^{-1}}$, and are consistent with having previously been companion WDs in pre-SN~Ia systems.  However, although the radial velocity of one of the stars is $>\unit[1000]{km \, s^{-1}}$, the radial velocities of the other two stars are puzzlingly consistent with 0.  The combined five-parameter astrometric solutions from \emph{Gaia} and radial velocities from follow-up spectra yield tentative 6D confirmation of the D$^6$ scenario.  The past position of one of these stars places it within a faint, old SN remnant, further strengthening the interpretation of these candidates as hypervelocity runaways from binary systems that underwent SNe~Ia.

\end{abstract}



\section{Introduction}
\label{sec:intro}

Type Ia supernovae (SNe~Ia) are one of the most common types of SNe in the local Universe.  They are best known for their utility as cosmological standardizable candles \citep{ries98,perl99} and also play a crucial role in galactic chemical evolution \citep{tww95}.  There is general agreement that the exploding star is a carbon/oxygen white dwarf (C/O WD) and that a companion star triggers runaway nuclear fusion in the WD, leading to a SN~Ia powered by the decay of radioactive $^{56}$Ni \citep{pank62a,cm69,maoz14a}.  However, despite decades of focused effort, there is no consensus regarding the nature of the companion or the mechanism by which the WD explodes, or even more fundamentally, whether one or multiple progenitor scenarios are responsible.

In ``single-degenerate'' scenarios, the companion is a non-degenerate hydrogen- or helium-burning star, while in ``double-degenerate'' scenarios, the companion is another WD.  These companions may trigger an explosion in the primary WD in a variety of ways that, in some cases, can be shared among different companion types. For example, models in which the growth of the primary WD leads to the ignition of convective carbon burning, causing a deflagration, detonation, and subsequent SN~Ia, have been proposed for hydrogen-rich single-degenerate donors \citep{wi73,nomo82a}, helium-rich single degenerates \citep{yoon03a}, and double-degenerate merger remnants \citep{it84,webb84}.  The double-detonation mechanism, in which a helium shell detonation sets off a carbon core detonation \citep{taam80b,livn90,shen14a}, has been proposed for helium single degenerates \citep{nomo82b,wtw86} and for double degenerates that undergo either stable \citep{bild07,fhr07,fink10,sb09b} or unstable mass transfer \citep{guil10,dan11,rask12,pakm13a,dan15a}.  Note that the secondary WDs in these double-degenerate double-detonation systems do not have to be helium core WDs, because C/O WDs are born with significant surface helium layers.

In all progenitor scenarios except for the subclasses of double-degenerate scenarios in which the companion WD is completely destroyed \citep{it84,webb84,pakm12b,papi15a}, the companion star will survive the explosion of the primary WD.  This surviving companion will fly away from the site of the explosion with the orbital velocity it had prior to the explosion.  The impact of the SN ejecta will also strip material from the companion, deposit shock energy, and possibly pollute the remaining surface layers.  These effects lead to observable peculiarities of varying degree, depending on the nature of the companion \citep{mbf00,pakm08a,pan13a,shap13a,shen17a}.

Searches for surviving companions within the remnants of historical SNe have predominantly focused on the relatively slow ejection velocities and small search radii implied by single-degenerate scenarios (e.g., \citealt{ruiz04b,sp12,kerz14c}).  These studies have failed to conclusively discover any surviving companions, which may not be surprising given other mounting evidence that single-degenerate scenarios cannot be responsible for the bulk of SNe~Ia (e.g., \citealt{leon07,kase10,gb10,li11,bloo12,olli15a,magu16a,wood17a}).

Only one study by \cite{kerz18a} has covered a large enough search region to probe the $\unit[1000-2500]{km \, s^{-1}}$ runaway velocities expected for surviving WD companions.  In their examination of the remnant of SN 1006, no bright, blue sources resembling those predicted by \cite{shen17a} were found within $8.5'$ of the remnant's center.  However, if the companion WD has cooled significantly since the SN, it may appear as a more typical-looking WD, which would be difficult to distinguish among the other normal stars within SN 1006's remnant.  Furthermore, if significant iron-line blanketing occurs, or if the SN otherwise dramatically alters the companion's appearance, the surviving WD would not be as blue as predicted by \cite{shen17a}; such redder sources do exist within the remnant but have yet to be systematically followed up.

In this work, we use \emph{Gaia}'s second data release (DR2; \citealt{gaia16a,gaia18a}) to perform an all-sky search for hypervelocity surviving companion WDs.  In Section \ref{sec:motandprop}, we make predictions for the expected state and number of surviving WDs that may be detected by \emph{Gaia}.  In Section \ref{sec:gaia}, we describe our search for hypervelocity WDs in DR2.  We detail our findings and our follow-up efforts in Section \ref{sec:followup}, and we conclude in Section \ref{sec:conc}.


\section{Motivation for and properties of a surviving companion WD}
\label{sec:motandprop}

In this section, we motivate the possibility that a companion WD might survive the SN~Ia and describe its expected properties following the explosion.  We also calculate the expected number of such sources in \emph{Gaia} DR2 under the assumption that all SNe~Ia yield a surviving companion WD.  We note here that we distinguish between companion WDs that survive normal SNe~Ia and the kicked bound WD remnants of explosions that may lead to the peculiar class of SNe~Iax \citep{jord12a,fole13a,long14a,fink14a}.  While extremely interesting, the surviving WD primaries of SN~Iax explosions are not expected to reach velocities higher than $\unit[1000]{km \, s^{-1}}$ and are not the subject of discussion in this section.  We also draw a distinction here between the hypervelocity ($> \unit[1000]{km \, s^{-1}}$) WDs we discuss in this work and the $\ll \unit[1000]{km \, s^{-1}}$ WDs predicted in previous studies that result from the evolution of non-degenerate survivors of single-degenerate SN~Ia progenitor scenarios \citep{hans03a,just09a}.


\subsection{Motivation}

Mass transfer between two WDs can be dynamically stable or unstable, depending on the donor's response to mass loss, whether or not the mass transfer is conservative, and the degree to which the angular momentum of the transferred mass can be converted back into orbital angular momentum \citep{mns04}.  For large enough mass ratios $\gtrsim 0.2$, the accretion stream directly impacts the more massive WD, and no accretion disk is formed, which implies inefficient angular momentum transfer to the orbit.  Furthermore, the mass transfer can be super-Eddington and thus non-conservative.  Both of these effects destabilize the binary, and thus most double WD binaries are expected to undergo dynamically unstable mass transfer.

Double WD binaries with extreme mass ratios $\lesssim 0.2$ can have sub-Eddington accretion rates and form accretion disks, both of which help to stabilize mass transfer.  However, \cite{shen15a} points out that even these systems can be driven to unstable mass transfer.  The accumulation of the donor's hydrogen- and helium-rich layers on the accretor lead to classical-nova-like events in which the accreted shell expands relatively slowly due to thermonuclear burning.  Dynamical friction between the donor and the expanding envelope causes the binary separation to decrease, which increases the mass transfer rate into the super-Eddington regime and destabilizes even these extreme mass ratio binaries.  This theoretical possibility is borne out by the relatively low birth rate of AM CVn systems \citep{brow16b}, providing evidence that all double WD systems eventually undergo runaway mass transfer.

In the original double-degenerate scenario \citep{it84,webb84}, such double WD systems undergoing dynamically unstable mass transfer coalesce to form a single merger remnant that explodes as a SN~Ia after $\sim \unit[10^4]{yr}$.  However, recent studies have suggested that the SN~Ia can be triggered during the coalescence itself due to the presence of helium in the surface layers of the companion WD, which initiates the SN~Ia explosion via the double-detonation mechanism \citep{guil10,dan11,rask12,pakm13a,dan15a}.  We term this combination of explosion mechanism, companion star, and mode of mass transfer the ``dynamically driven double-degenerate double-detonation'' (D$^6$) scenario.  In fact, when all appropriate nuclear reactions are accounted for \citep{shen14b}, helium shell detonations, and subsequent carbon core detonations, may even occur during the relatively quiescent initial phases of dynamical mass transfer, before the complete tidal disruption of the companion WD \citep{pakm13a,shen17a}.  Thus, there is the exciting possibility that the D$^6$ scenario yields a smoking gun in the form of a surviving companion WD.

While theoretical confirmation of the existence of a surviving companion WD awaits future detailed simulations, several observables suggest  it is a necessary outcome if double WD binaries make up the bulk of SN~Ia progenitors.  If fully disrupted, the companion WD would form a thick torus surrounding the primary WD \citep{ggi04,dan14a}, which would impart significant asymmetry to the ejecta when the explosion occurred \citep{rask14a}.  However, such asymmetry is at odds with the low levels of polarization measured in SNe~Ia \citep{wang08a,bull16a}.  Furthermore, much of the disrupted WD material would remain at low velocities throughout the evolution of the SN \citep{pakm12b}.  In the late-time nebular phase, when the SN ejecta becomes optically thin, oxygen in the disrupted WD material may become visible as strong, narrow emission features, but these are almost never detected (although see \citealt{krom13a} and \citealt{taub13a} for the case of SN 2010lp).  Given these observational constraints and theoretical motivation, we proceed under the presumption that most, if not all, SNe~Ia leave an intact companion WD.


\subsection{Luminosity}
\label{sec:lum}

An estimate of the number of runaway WDs that \emph{Gaia} will detect requires knowledge of their brightness following the SN~Ia explosion.  As a lower limit, we can assume that the evolution up to the explosion and its aftermath have no effect on the companion, so that its luminosity is the same as that of a WD that has cooled in isolation since its birth.  Using publicly available DB WD cooling tables\footnote{\url{http://www.astro.umontreal.ca/~bergeron/CoolingModels}} \citep{holb06a,kowa06a,trem11a,berg11a}, we find that a $0.6 \, (1.0) \msol$ WD that has cooled for $\unit[1]{Gyr}$ since its birth has an absolute visual magnitude of $ 12.9 \, (12.8)$.  However, this declines to $ 14.3 \, (14.0) $ by $\unit[3]{Gyr}$.  Given \emph{Gaia}'s magnitude limit of $ 21$, this lower limit to the luminosity implies a $3 \, {\rm Gyr}$-old $0.6 \msol$ WD can only be seen out to $\unit[200]{pc}$, while a $1 \, {\rm Gyr}$-old $1.0 \msol$  WD will be detected out to $\unit[400]{pc}$.

However, there are several processes that will increase the luminosity of the companion above this lower limit prior to and just after the SN~Ia explosion.  The most important of these is likely tidal heating.  As the orbit of the two WDs decays due to gravitational wave emission, the binary separation shrinks and the less massive companion begins to feel the tidal field of the primary.  If the companion becomes and remains tidally locked to the primary, viscous dissipation yields a luminosity at the onset of mass transfer ranging from $0.15-1000 \, L_\odot$, depending on the component masses \citep{iben98a}, far higher than that of an old, isolated WD.

These values represent upper limits to the effects of tidal heating, as the companion will not be able to maintain complete synchronicity with the orbit.  Recent work has found that tidal effects are dominated by the excitation of gravity waves within the companion, which deposit their energy and angular momentum near the surface of the WD \citep{fl11,burk13a}.  These surface layers can be kept near synchronous rotation all the way to the onset of mass transfer, but since the energy is not deposited deep within the core, the timescale for this excess heat to be radiated away is shorter than the typical $>\unit[10^6]{yr}$ between the explosion and the present day for nearby runaway WDs (Section \ref{sec:numest}), and thus the WD would cool and approach the luminosity of a dim, isolated WD.

However, as argued by \cite{burk13a}, the strength of the WD's fossil magnetic field should be more than adequate to maintain solid body rotation between the outer layers, where angular momentum is deposited, and the core, especially if the field is wound up during the evolution towards solid body rotation.  As the interior of the WD is spun up, it will also be heated by small scale turbulence at a comparable level to the dissipation necessary to maintain complete synchronicity with the orbit.  In this case, the physical picture approaches that assumed by \cite{iben98a}, and the luminosities at the point of Roche lobe overflow (RLOF) will approach their values.  Such luminosities $ \geq 0.1 \, L_\odot$ will be visible to $\unit[1]{kpc}$, a distance at which \emph{Gaia}'s faint end parallax errors become the primary limitation (Section \ref{sec:gaia}).  Thus, when calculating the expected number of runaway WDs \emph{Gaia} will find in Section \ref{sec:numest}, we limit our search volume to a sphere around the Sun with a radius of $\unit[1]{kpc}$.

In addition to tidal heating, there are several other possible mechanisms that may change the appearance of a surviving companion WD, all of which occur after the explosion of the primary.  The first is due to the impact of the SN ejecta on the companion, which will deposit shock energy and may also ablate some material from the surface.  While similar processes have been modeled for single-degenerate companions (e.g., \citealt{mbf00,pakm08a,liu12a,pan14a}), detailed calculations of the post-impact state of a surviving WD have not yet been performed, so quantitative predictions cannot be made (but see \citealt{papi15a} for a preliminary investigation of some of these effects).  However, given the $>\unit[10^6]{yr}$ average delay between the SN and the present day for local runaway WDs (Section \ref{sec:numest}) and the much shorter thermal time at the relatively shallow depths where this shock energy should be deposited, it seems unlikely that these effects will still be observable for all but the youngest WDs within historical SN remnants.

A second mechanism that may heat the surviving WD is due to the rapid expansion of the exploding primary, which causes the tidal field felt by the companion WD to change abruptly.  Dissipation during the subsequent relaxation to a new hydrostatic equilibrium will deposit heat throughout the WD.  However, as in the first case, most of the tidal deformation occurs near the surface of the WD, so that most of the dissipation will also be preferentially located at shallow depths where the thermal timescale is relatively short.  As above, given the large average age of the local surviving WDs, we expect this excess heat to be negligible except for ex-companions to historical SNe.

A final mechanism to increase the surviving companion's post-SN luminosity concerns the capture and accretion of $^{56}$Ni from the SN ejecta by the surviving companion.  As discussed in \cite{shen17a}, much of the high-entropy $^{56}$Ni remains fully ionized as it settles onto the WD's surface and cannot decay via standard bound electron captures until it cools.  Thus, the radioactive decay of this accreted $^{56}$Ni can keep the companion WD relatively bright centuries after the SN has faded.

As mentioned above, detailed calculations of the SN ejecta's interaction with the surviving WD have not yet been performed.  This means that stellar evolution calculations, like those in \cite{shen17a}, are forced to rely on simple estimates for the amount of radioactive material captured and its initial thermal state.  As such, accurate quantitative predictions cannot yet be made.  However, similarly to shock heating by the SN ejecta and tidal relaxation, it is likely that the delayed radioactive decays only affect the outer layers of the companion WD, limiting the time when they can contribute to the luminosity and alter the colors of the WD to centuries.  These effects are important for runaway WDs from historical SNe, such as Tycho, Kepler, SN 185, and SN 1006 \citep{kerz18a}, but may be negligible for the much older runaway WDs that should form the bulk of our candidates.  For most of the hypervelocity WDs  \emph{Gaia} is likely to detect, we expect that the energy deposited much deeper in the WD's interior from tidal heating prior to the SN~Ia will likely determine its present-day luminosity.


\subsection{Surface abundances}
\label{sec:abun}

As discussed in \cite{shen17a}, the surviving WD will capture some of the lowest velocity SN ejecta, primarily composed of $^{56}$Ni.  The energy released by the slowly decaying $^{56}$Ni blows a wind from the WD's surface, ejecting much of the accreted SN ejecta, but some of the material should remain bound and might be detectable with follow-up high resolution spectra.  Unfortunately, as before, the lack of relevant hydrodynamic simulations makes accurate predictions of the surface abundances of a surviving WD difficult.

A surviving WD companion should be hydrogen-free, because the hydrogen layer that most WDs are born with will have been transferred to the primary and ejected from the system in the $\sim \unit[1000]{yr}$ prior to the SN \citep{kbs12,shen13a,shen15a}.  Given the long cooling timescales of the expected \emph{Gaia} WDs within $\unit[1]{kpc}$, sedimentation may cause heavy metals to sink \citep{paqu86b,dupu92a}, leaving only helium or carbon and oxygen in the atmosphere, depending on the initial composition of the companion, but anomalous abundances could potentially still be observable for the young runaway WDs from Tycho, Kepler, SN 185, and SN 1006.  However, this statement depends on the existence and depth of a surface convection zone, the thermal profile, and, if the surface layers remain $\gtrsim \unit[2\E{4}]{K}$, the competing effect of radiative levitation \citep{chay95a,chay95b}.  We regard the expected surface abundances to be uncertain and thus one of the motivations for follow-up spectroscopy.


\subsection{Velocity}

Given the uncertainties in the observational characteristics discussed in the previous sections, the clearest and most obvious distinguishing feature of a surviving companion WD is its hypervelocity.  The orbital velocity of a companion star during Roche lobe overflow (RLOF), which will be its runaway velocity once the primary WD explodes,\footnote{In principle,  the runaway velocity is somewhat smaller than the pre-SN orbital velocity due to the portion of the exploding WD that remains at velocities $\lesssim v_{\rm runaway}$.  However, in practice, this mass is always $<0.05 \msol$ \citep{shen18a} and will have a negligible effect on the runaway velocity.} is a function of its mass, $M_2$, radius, $R_2$, and the mass of the exploding WD, $M_1$:
\begin{align}
	v_{\rm runaway} = \sqrt{ \frac{ GM_1^2 }{ \left( M_1+M_2 \right) a } } ,
\end{align}
where the binary separation during RLOF, $a$, is approximated by \cite{eggl83} as
\begin{align}
	a = \frac{R_2}{0.49} \left\{ 0.6 + \left( \frac{M_1}{M_2} \right)^{2/3} \ln \left[ 1+ \left( \frac{M_2}{M_1} \right)^{1/3} \right] \right\} .
\end{align}

\begin{figure}
  \centering
  \includegraphics[width=\columnwidth]{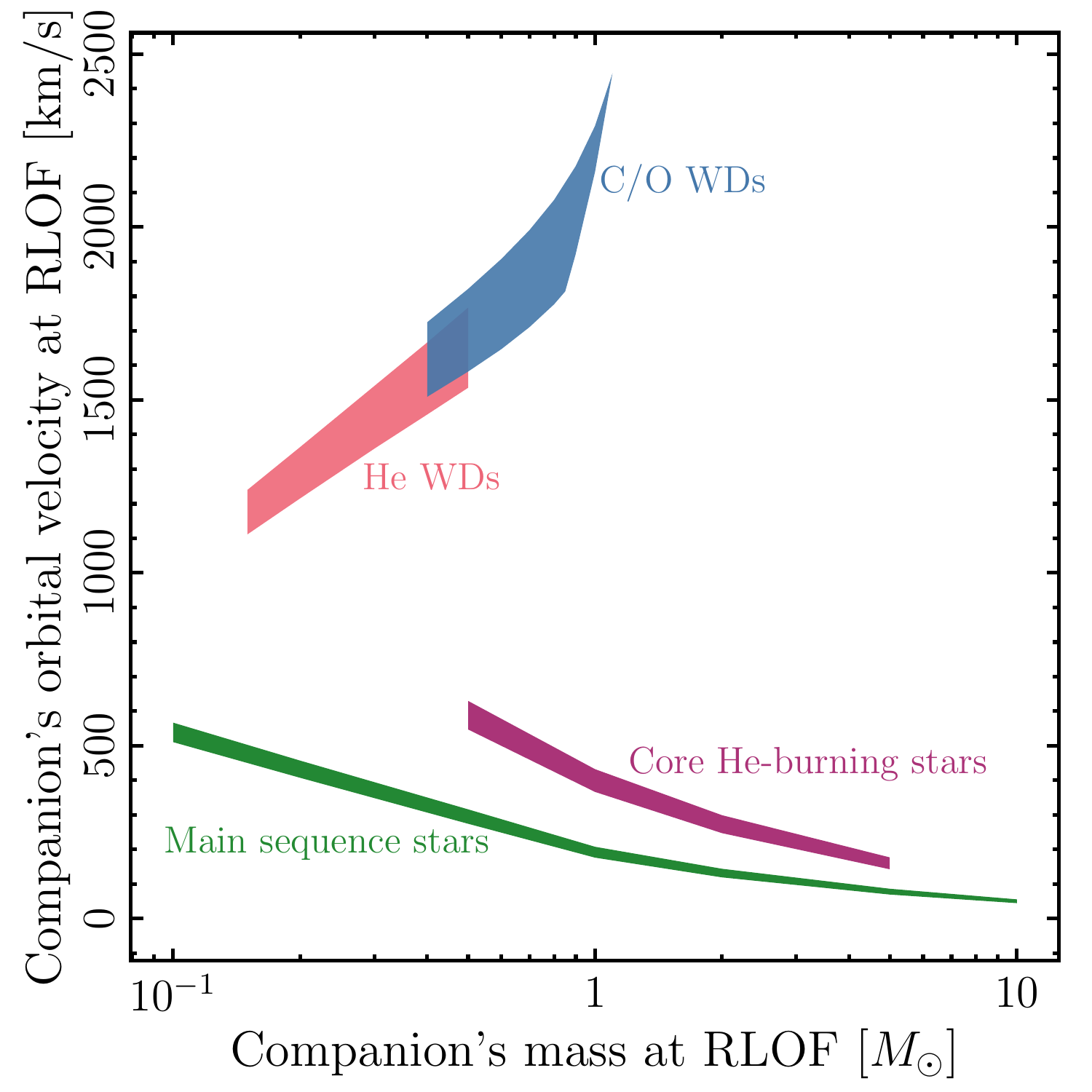}
  \caption{The companion's orbital velocity vs.\ mass at RLOF.  The upper boundary of each region corresponds to a $1.1 \msol$ primary WD; the lower corresponds to a $0.85 \msol$ primary.}
  \label{fig:mvsv}
\end{figure}

We calculate simple mass-radius relationships for isolated helium and C/O WDs, as well as for hydrogen- and helium-burning non-degenerate companions with the stellar evolution code \texttt{MESA} \citep{paxt11,paxt13,paxt15a,paxt18a}.  The resulting runaway velocities vs.\ companion masses are shown in Figure \ref{fig:mvsv}.  We assume the mass of the exploding WD ranges from $0.85 \msol$ (lower boundary of each region) to $1.1 \msol$ (upper boundary), as motivated by WD detonation calculations \citep{sim10,blon17a,shen18a}.

Both non-degenerate and degenerate SN~Ia companions undergo processes that increase their radii as compared to isolated stars.  As discussed in Section \ref{sec:lum}, tidal effects in the D$^6$ scenario lead to dissipation throughout the companion WD, including near its surface where it may cause radial expansion.  In single-degenerate scenarios, the long pre-explosion phase of mass transfer pushes the donor out of thermal equilibrium and leads to ``thermal bloating'' \citep{knig11a}.  Thus, the actual companion radii when the SN~Ia explosion occurs may be slightly larger than the radii of isolated stars we find with \texttt{MESA}.  However, since the companion's orbital velocity scales as the square root of its radius, a 10\% increase in the radius, e.g., only corresponds to a 5\% decrease in the velocities shown in Figure \ref{fig:mvsv}.

It is clear that the runaway velocities of surviving companion WDs will be markedly different from those of any other possible companions.  A fiducial $0.6 \msol$ companion to a $1.0 \msol$ primary WD will have a runaway velocity of $\unit[1800]{km \, s^{-1}}$, while the companion of a near-equal mass ratio $1.0+1.0 \msol$ binary will have a velocity of $\unit[2200]{km \, s^{-1}}$.  At a distance of $\unit[1]{kpc}$, these velocities translate to proper motions of $0.4-0.5'' \, {\rm yr^{-1}}$ if the velocities are in the plane of the sky, easily detectable by \emph{Gaia} and below the very high proper motions $\geq 0.6'' \, {\rm yr^{-1}}$ where completeness becomes an issue \citep{gaia16a,gaia18a}.


\subsection{Estimated number of runaway WDs}
\label{sec:numest}

As discussed in the previous sections, \emph{Gaia} will only be able to detect surviving companion WDs that are relatively nearby ($ \leq \unit[1]{kpc}$) or that were reheated by historical SNe and have not yet dimmed significantly.  The SN remnant most likely to host a surviving WD observable with \emph{Gaia} is SN 1006, due to its proximity ($\simeq \unit[2.2]{kpc}$), low extinction, and lack of crowding \citep{wink03a}.  Although \cite{kerz18a} did not detect any young WDs on the WD cooling track within this remnant, it is possible that the accreted iron-group elements have shifted the spectral energy distribution redward so that the surviving WD no longer sits on the WD cooling track.  Indeed, some sources $ \lesssim \unit[1]{mag}$ redder than the cooling track do exist within the remnant; we will specifically target these and other sources within SN 1006's remnant in Section \ref{sec:gaia}, as well as performing searches within the remnants of SN 185 (RCW 86) and Tycho's and Kepler's SNe.

In order to estimate the number of runaway WDs unassociated with historical SNe within $\unit[1]{kpc}$ of the Sun, we require a model for the stellar density as a function of position within the Milky Way.  We assume all the stars reside in the thin disk, with an exponential scale length of $\unit[2.6]{kpc}$, an exponential scale height of $\unit[300]{pc}$, and a normalization that yields a total stellar mass of $5\E{10} \msol$ \citep{blan16a}.  We take the Sun's Galactocentric distance to be $\unit[8.2]{kpc}$, its height from the midplane to be $\unit[25]{pc}$, and the Milky Way's specific SN~Ia rate to be $10^{-13} \, {\rm SNe} \, {\rm yr^{-1}} \, M_\odot^{-1}$ \citep{li11c}.

With these values and additional assumptions that the Galactic potential is negligible and that the runaway WDs all have the same velocity of $1800 \, (2200) \, {\rm km \, s^{-1}}$,  we find the expected number of nearby runaway WDs to have a Poissonian distribution centered at $28 \, (23)$, with a 95\% confidence interval of $17-39 \, (14-33)$.  Of these, $22 \, (18)$ will be more than $10^\circ$ off the plane of the Milky Way and thus will not be subject to significant extinction or crowding.  These numbers are significantly higher than the $\lesssim 1$ expected hypervelocity stars ejected from the Galactic center currently passing within $\unit[1]{kpc}$ \citep{hill88a,brow15a}, so we do not expect contamination to be a concern. Even if the rate of hypervelocity ejections from the Galactic center is much higher than previously thought, luminosities, colors, spectra, and whether or not their orbits intersect the Galactic center should allow us to easily differentiate between runaway WDs and hypervelocity stars.

\begin{figure}
  \centering
  \includegraphics[width=\columnwidth]{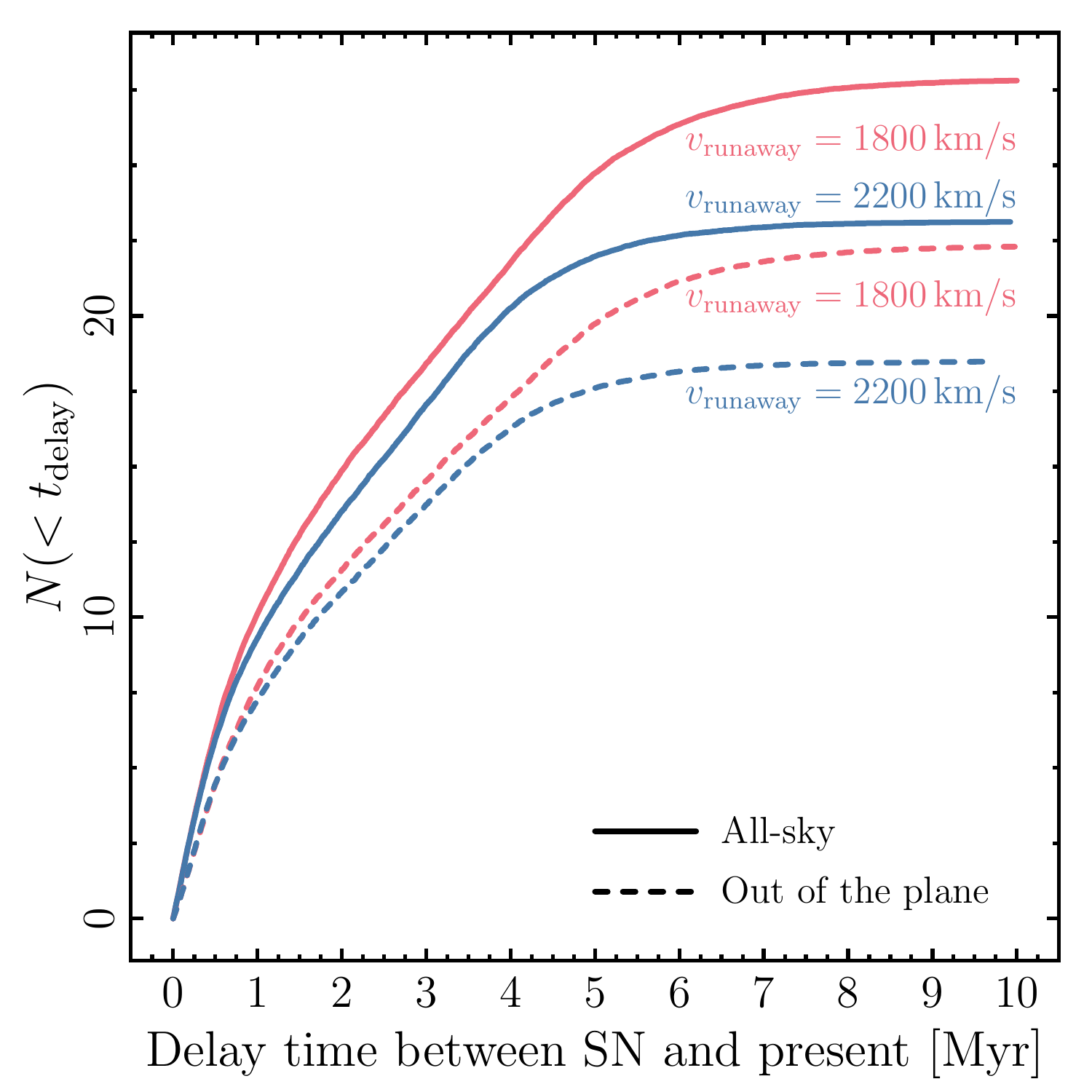}
  \caption{Cumulative distribution functions of the time between the SN~Ia event that ejected the runaway WD and the present day.  Solid lines represent CDFs for the whole volume within $\unit[1]{kpc}$; dashed lines represent runaway WDs $>10^\circ$ off the Galactic plane.}
  \label{fig:agecdf}
\end{figure}

The cumulative distribution function (CDF) of delay times, $t_{\rm delay}$, between the SN explosion and the present for these local runaway WDs is shown in Figure \ref{fig:agecdf}.  The average elapsed time for runaway WDs moving at $1800 \, (2200) \, {\rm km \, s^{-1}}$ is $2.3\E{6} \, (1.5\E{6}) \, {\rm yr}$.  As discussed in Section \ref{sec:lum}, this value is longer than the thermal timescale at the depth where gravity waves dissipate tidal energy, but it is shorter than the thermal timescale from the center of the WD to the surface.  Thus, as long as significant tidal heating is deposited deep within the companion, these runaways should be visible to at least $\unit[1]{kpc}$ for \emph{Gaia}'s limiting magnitude of $21$.

\begin{figure}
  \centering
  \includegraphics[width=\columnwidth]{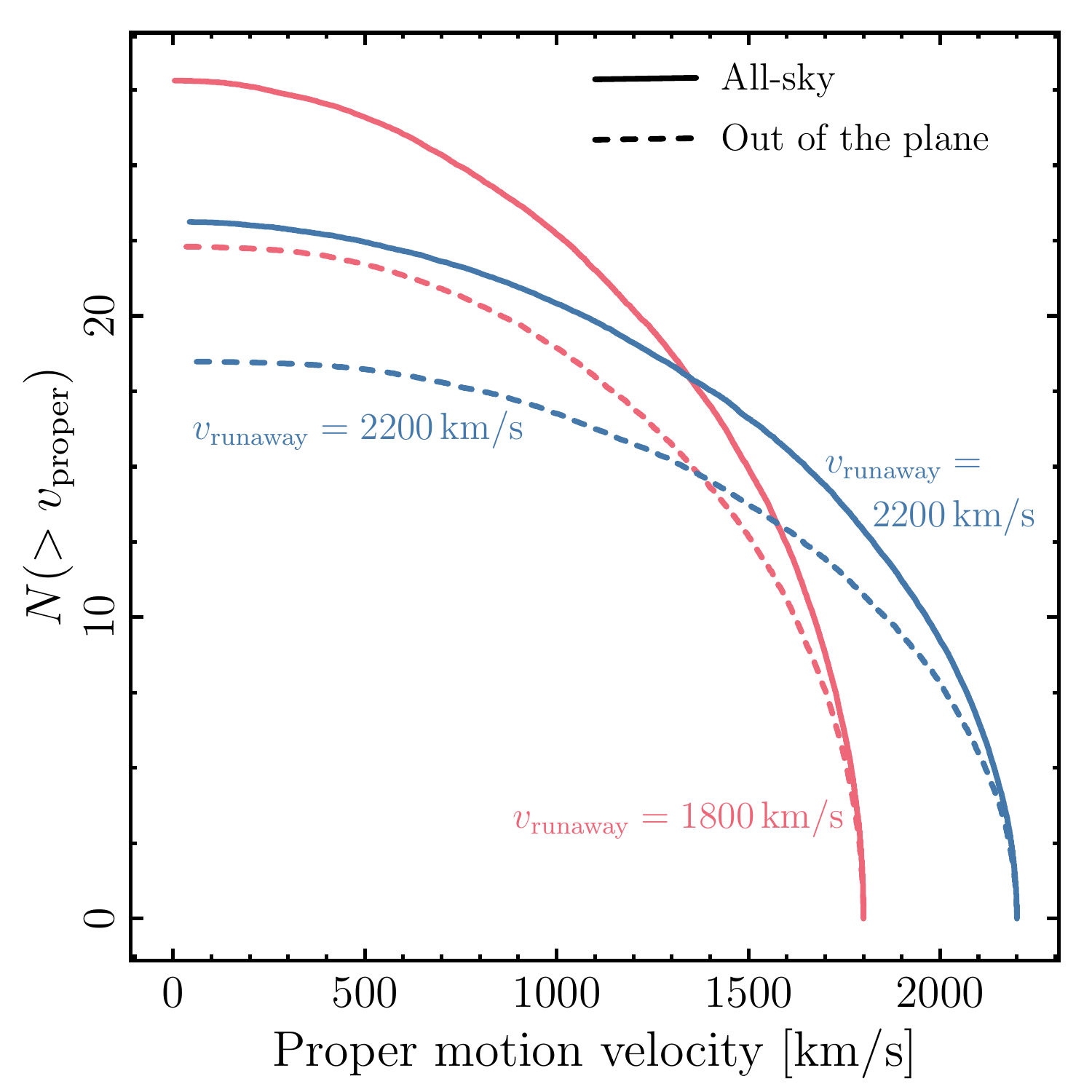}
  \caption{Cumulative distribution functions of the runaway WD velocity in the plane of the sky.  Solid lines represent CDFs for the whole volume within $\unit[1]{kpc}$; dashed lines represent runaway WDs $>10^\circ$ off the Galactic plane.}
  \label{fig:vcdf}
\end{figure}

A similar CDF of proper motion velocities, $v_{\rm proper}$, is shown in Figure \ref{fig:vcdf}.  Geometric effects imply that the velocity of the WDs in the plane of the sky will be smaller than their total velocities, but over $80\%$ of the nearby runaway WDs will still have  $v_{\rm proper} >\unit[1000]{km \, s^{-1}}$.  At a distance of $\unit[1]{kpc}$, $\unit[1000]{km \, s^{-1}}$ translates to a proper motion of $0.2'' \, {\rm yr^{-1}}$, which is much larger than the expected proper motion errors  at \emph{Gaia}'s magnitude limit.  Furthermore, while $\sim 20\%$ of stars with proper motions $ \geq 0.6'' \, {\rm yr}$ may be missing from DR2 \citep{gaia16a}, only the $\sim 1$ runaway WD expected to be closer than $ \unit[300]{pc}$ will be affected.


\section{Candidate selection in \emph{Gaia} DR2}
\label{sec:gaia}

\emph{Gaia} was launched in December 2013 and has been recording the precise astrometry of billions of stars since July 2014 \citep{gaia16a}.  When its nominal five-year science mission is complete, it will have measured parallaxes, $\varpi$, and proper motions, $\mu$, of most of the stars brighter than $G \simeq 21$.  In this section, we outline our search strategy and describe the hypervelocity WD candidates found in \emph{Gaia}'s second data release (DR2), which occurred on 25 April 2018 and provided astrometric parameters of $\simeq 1.3\E{9}$ stars.


\subsection{Search strategy}

The relatively small number ($\sim 30$) of expected local hypervelocity runaway WDs within $\unit[1]{kpc}$ is equivalent to the number of $6$-$\sigma$ outliers in a normally distributed set of samples as large as \emph{Gaia}'s dataset.  We must therefore exercise caution in our search strategy in order to avoid being overwhelmed by false positives.

We begin by restricting the $1.3\E{9}$ sources to those with proper motions above a conservative limit equivalent to $\unit[1000]{km \, s^{-1}}$ at $\unit[1]{kpc}$.  For such proper motions $\mu \geq \unit[211]{mas \, yr^{-1}}$, the fractional errors are $<0.01$, so we neglect them for simplicity.  A $\varpi > 3 \sigma_\varpi$ cut is also applied, although two sources with $\varpi < 3 \sigma_\varpi$  (O1 and O3 in Table \ref{tab:cand}) were followed up prior to the implementation of this cut.  The sources are then rank-ordered by $v_{3 \sigma} \equiv \mu/(\varpi + 3 \sigma_\varpi)$, where $\sigma_\varpi$ is the parallax error.  By including the $3 \sigma_\varpi$ term, we are effectively calculating the proper motion velocity using the $3$-$\sigma_\varpi$ upper bound for the parallax.

Since the fractional parallax errors can be $\sim 1$, we apply a Bayesian framework to the top 500 entries to better quantify the posterior probability that the candidates are actually hypervelocity WDs. We use an exponentially decreasing space density distance prior \citep{astr16a},\footnote{Other priors, such as a uniform distance prior, were also tried, but they do not significantly affect the ordering of the candidate list.} assume the likelihoods of the parallaxes are normally distributed, and calculate posteriors
\begin{align}
	P(r | \varpi, \sigma_\varpi) = \frac{ P( \varpi | r, \sigma_\varpi) P(r)}{ \int P( \varpi | r, \sigma_\varpi) P(r) } .
\label{eqn:post}
\end{align}
Finally, we integrate the posterior with appropriate limits to find the probability, $P_{1000}$, that the proper motion velocity is larger than $\unit[1000]{km \, s^{-1}}$, and the probability, $P_{1000-3000}$, that it is bounded between 1000 and $\unit[3000]{km \, s^{-1}}$.

Note that the calculated probabilities, $P_{1000}$ and $P_{1000-3000}$, should not be taken literally, as they may be strongly influenced by the non-Gaussianity of the parallax distribution's tails above $\sim 4$-$\sigma_\varpi$ \citep{luri18a}.  We merely use these probabilities as a guide for ranking our sources for follow-up.

We slightly alter our strategy to additionally search for surviving WDs within the four Galactic remnants of suspected SNe~Ia: the remnant of SN 185 (RCW 86), at a distance of $\unit[2.0-3.0]{kpc}$ \citep{held13a}, the remnant of SN 1006 ($\unit[2.1-2.3]{kpc}$; \citealt{wink03a}), Tycho's SN remnant ($\unit[3-5]{kpc}$; \citealt{haya10a}), and Kepler's SN remnant ($\unit[3-6]{kpc}$; \citealt{sank05a,chio12a}).  We apply the same search strategy as above, but we relax our proper motion cuts to the equivalents of $\unit[1000]{km \, s^{-1}}$ at the upper limits for the remnants' distances.  We also restrict our search regions to circles around the geometric centers of the remnants with radii corresponding to proper motion velocities of $\unit[4000]{km \, s^{-1}}$ at the lower limits for the distances.  This high proper motion velocity accounts for the combination of the surviving companion WD's velocity and the initial velocity of the exploding WD and its remnant.


\subsection{List of hypervelocity candidates}

The \emph{Gaia} source IDs of seven of the candidates that we were able to follow up with ground-based instruments are shown in Table \ref{tab:cand}, along with their nicknames, which we will use hereafter, astrometric parameters, values for $v_{3 \sigma}$, and probabilities, $P_{1000}$ and $P_{1000-3000}$.  The associated photometry from \emph{Gaia}, PS1 \citep{cham16a}, and Skymapper \citep{wolf18a} and the telescopes used to follow up the sources are shown in Table \ref{tab:phot}, along with comments about the individual stars.  None of the searches within the Galactic SN~Ia remnants revealed any obvious hypervelocity candidates.  

\begin{deluxetable*}{cc|cccccccc}
\tablecaption{List of candidate hypervelocity WDs followed up in this work \label{tab:cand}}
\tablehead{
\colhead{\emph{Gaia} DR2 ID} & \colhead{Nickname} & \colhead{R.A.} & \colhead{Dec.} & \colhead{Parallax} & \colhead{$\mu_{\alpha^*}$\tablenotemark{a}} & \colhead{$\mu_\delta$\tablenotemark{b}} & $v_{3 \sigma}$\tablenotemark{c}& \colhead{$P_{1000}$\tablenotemark{d}}  & \colhead{$P_{1000-3000}$\tablenotemark{e}} \\
\colhead{} & \colhead{} & \colhead{(degrees)} & \colhead{(degrees)} & \colhead{(mas)}  & \colhead{($\unit[]{mas \, yr^{-1}}$)} & \colhead{($\unit[]{mas \, yr^{-1}}$)} & \colhead{$\unit[]{km \, s^{-1}}$} &  \colhead{} & \colhead{} }
\startdata
5805243926609660032 & D6-1 & 249.3819752 & -74.3434986 & $0.471 \pm 0.102 $ & $-80.3 \pm 0.1$ & $-195.9 \pm 0.2$ & 1293 & 1.00 & 0.79 \\
1798008584396457088 & D6-2 & 324.6124885 & 25.3737115 & $1.052 \pm 0.109$ & $98.4 \pm 0.2$ & $240.4 \pm 0.2$ & 894 & 0.98 & 0.98  \\
2156908318076164224 & D6-3 & 283.0078540 & 62.0361675 & $0.427 \pm 0.126$ & $9.0 \pm 0.2$ & $211.5 \pm 0.3$ & 1247 & 1.00 & 0.57  \\
2050179518946705152 & O1 & 291.3306894 & 36.4500600 & $0.237 \pm 0.317$ & $-137.6 \pm 0.6 $ & $-214.7 \pm 0.6 $ & 1018 & 1.00 & 0.17  \\
4396109004117478656 & O2 & 237.5476987 & -7.8881665 & $2.186 \pm 0.225$ & $-359.7 \pm 0.5 $ & $-228.4 \pm  0.3 $ & 706 & 0.37 & 0.37  \\
5884527618445501056 & O3 & 238.9823874 & -55.4937974 & $0.360 \pm 0.571$ & $90.0 \pm 0.8  $ & $-268.9 \pm 0.8  $ & 649 & 1.00 & 0.20 \\
1820931585123817728 & O4 & 296.3894478 & 17.2130727 & $0.574 \pm 0.076$ & $-82.4 \pm 0.1$ & $-149.5 \pm 0.1$ & 1010 & 1.00 & 1.00 \\
\enddata
\tablenotetext{a}{Heliocentric proper motion in right ascension.}
\tablenotetext{b}{Heliocentric proper motion in declination.}
\tablenotetext{c}{Heliocentric proper motion velocity calculated using the 3-$\sigma_\varpi$ upper bound of the parallax.}
\tablenotetext{d}{Probability that the proper motion velocity is larger than $\unit[1000]{km \, s^{-1}}$.}
\tablenotetext{e}{Probability that the proper motion velocity is bounded by 1000 and $\unit[3000]{km \, s^{-1}}$.}
\tablecomments{This is not a complete list of the most highly ranked hypervelocity WD candidates.  It only shows the stars that were followed up in this work.}
\end{deluxetable*}

\begin{deluxetable*}{c|ccccccccccccccc}
\tablecaption{Photometry and follow-up of hypervelocity  WD candidates \label{tab:phot}}
\tablehead{
\colhead{Nickname} & \multicolumn{3}{c}{\emph{Gaia}}& \multicolumn{5}{c}{PS1} & \multicolumn{5}{c}{Skymapper} & \colhead{Telescope} & \colhead{Comments} \\
\colhead{} & \colhead{\emph{G}} & \colhead{\emph{G$_{BP}$}} & \colhead{\emph{G$_{RP}$}} & \colhead{\emph{g}} & \colhead{\emph{r}} & \colhead{\emph{i}} & \colhead{\emph{z}} & \colhead{\emph{y}} & \colhead{\emph{u}} & \colhead{\emph{g}} & \colhead{\emph{r}} & \colhead{\emph{i}} & \colhead{\emph{z}} & \colhead{} & \colhead{} }
\startdata
D6-1 & $17.4$ & $17.6$ & $ 17.1$ &  --- & --- & --- & --- & ---  & 18.4 & 17.6 & 17.4 & 17.4 & ---  & SALT & D$^6$ WD candidate\\
D6-2 & 17.0 & 17.1  & 16.7  & 17.1 & 17.0 & 17.1 & 17.2 & 17.2 & --- & --- & --- & --- & --- & NOT & D$^6$ WD candidate\\
D6-3 & 18.3 &18.4 & 18.0 & 18.6 & 18.3 & 18.3 & 18.4 & 18.5 & --- & --- & --- & --- & --- & NOT & D$^6$ WD candidate\\
O1 & 17.0 & 17.6 & 16.0 & 18.0 & 17.0 & 16.5 & 16.2 & 16.1 & --- & --- & --- & --- & --- & NOT &  Ordinary star \\
O2 & 18.0 & 18.8 & 17.1  & 19.1 & 19.1 & 18.1 & 17.5 & 17.2 & --- & --- & --- & --- & --- & Bok &  Ordinary star \\
O3 & 17.8 & --- & --- &  --- & --- & --- & --- & ---  & --- & 16.7 & 16.0 & --- & ---  & SALT &  Ordinary star \\
O4 & 16.0 & 16.5 & 15.2 &  --- & --- & --- & --- & ---   & --- & --- & --- & --- & --- & Shane & Ordinary star \\
\enddata
\end{deluxetable*}


\section{Analysis of the hypervelocity candidates}
\label{sec:followup}

The seven candidates listed in Tables \ref{tab:cand} and \ref{tab:phot} were followed up at the Bok telescope (2.3-m), the Nordic Optical Telescope (NOT, 2.5-m), the Shane telescope at the Lick Observatory (3.0-m), and the Southern African Large Telescope (SALT, 9.8-m).  Of these seven candidates, the spectra show that four (nicknamed O1-O4) are ordinary-looking hydrogen-rich stars whose true parallaxes are likely much larger than measured, implying they are nearby stars with more pedestrian proper motion velocities.  We do not discuss these stars here further.  We now turn to the three remaining candidates.


\subsection{Spectroscopic follow-up}

\begin{figure*}
  \centering
  \includegraphics[width=0.9\textwidth]{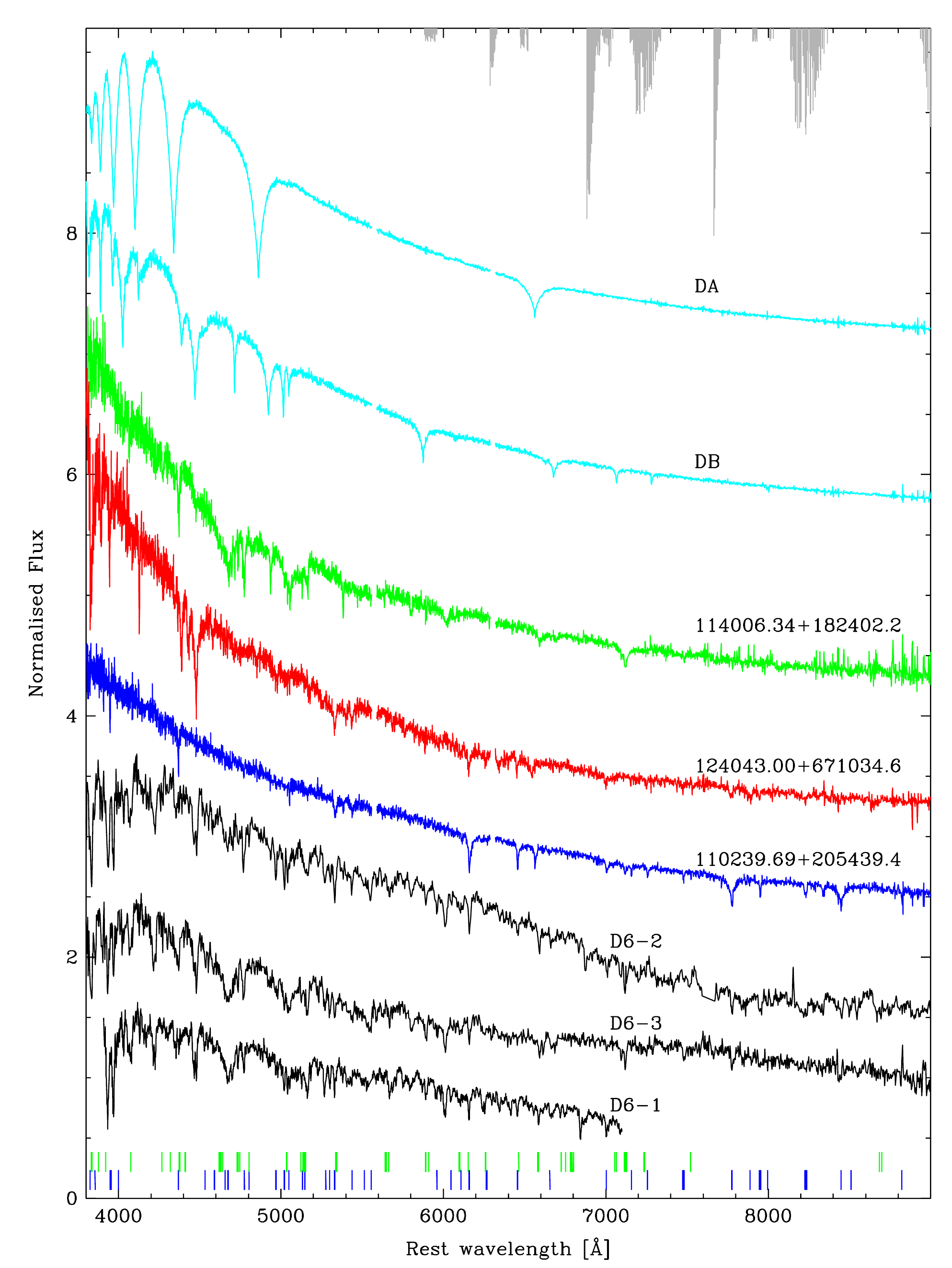}
  \caption{Identification spectra of the three runaway WD candidates (black) compared to SDSS spectra of WDs with large photospheric abundances of oxygen (blue), oxygen/magnesium/silicon (red) and carbon (green). The major lines of oxygen and carbon are indicated by blue and green tickmarks, respectively, near the bottom. Hydrogen and helium WD atmospheres (cyan) are clearly ruled out. Regions of the spectrum affected by telluric absorption are indicated in gray at the top of the figure.  The spectra of the three D$^6$ candidates are available as data behind the figure.}
  \label{fig:specWD}
\end{figure*}

The SALT data for D6-1 were obtained through Director's Discretionary Time (proposal 2017-2-DDT-005, PI: Jha) and made use of the Robert Stobie Spectrograph (RSS), with a $1.5''$ wide longslit and the PG0900 grating, resulting in a spectral resolution $\lambda/\Delta\lambda \approx 900$ over the wavelength range $\unit[392-713]{nm}$. The data were reduced with a custom pipeline that incorporates routines from PyRAF and PySALT \citep{craw10a}.

D6-2 and D6-3 were observed using the ALFOSC spectrograph on the Nordic Optical Telescope, located at Roque de Los Muchachos on La Palma.  All observations were taken using Gr4, which covers the region from $3400 - 9000 \, {\rm \AA}$ at low resolution, and a $1''$ slit oriented at the parallactic angle. Weather conditions were excellent, with seeing below 1\arcsec.  Spectra were reduced using the dedicated pipeline ALFOSCGUI. Overscan and bias subtraction and flat fielding were performed before 1D spectra were optimally extracted. The dispersion solution for the spectra was obtained from arc lamps taken with the same configuration as the science spectra at the start of the night. In addition, a linear wavelength shift was applied to each spectrum based on sky emission lines. Telluric absorptions have been corrected for using observations of a spectrophotometric standard. The resolution of the spectra (as measured from narrow sky lines in the spectra) was $\sim15 \, $\AA, while the S/N ratio for D6-2 and D6-3 was $\sim 25$.

Radial velocities (RVs) were measured via cross-correlation against the MILES library, comprising nearly 1000 stellar templates \citep{sanc06a,falc11a}. We used the \texttt{rvsao} \citep{kurt98a} implementation of the \cite{tonr79a} algorithm. We find over one hundred good matches in the template library for each of D6-1, D6-2, and D6-3, based on the reported $r$ statistic (ranging from $r =$ 4 to 8). We average the resulting RVs and take the scatter across templates to be indicative of the scale of potential systematic uncertainties. We note that cross-correlation among D6-1, D6-2, and D6-3 themselves results in a much higher $r \simeq 15$, showing that these spectra are much more similar to each other than they are to entries in the template library.  The resulting RV shifts are shown in Table \ref{tab:vel}.

Figure \ref{fig:specWD} shows the spectra of D6-1, D6-2, and D6-3.  An RV shift of $\unit[1200]{km \, s^{-1}}$ is applied to D6-1, while the spectra of D6-2 and D6-3 are unshifted.  The spectra contain a multitude of absorption features. No transitions of hydrogen or helium are detected, ruling out canonical atmosphere compositions \citep{klei13a}.  Comparison with unusual WDs identified by SDSS \citep{gaen10a,gent15a,kepl16a} clearly reveals strong features of carbon, oxygen, magnesium, and calcium in the three candidates.  Based on their $G_\textrm{BP}-G_\textrm{RP}$ colors, which are slightly redder than those of GD\,492, SDSS\,1102+2054, and SDSS\,1140+1824 (see Section \ref{sec:cmd}), we estimate effective temperatures of $\simeq 8000$\,K. Determining accurate atmospheric parameters for stars with such unusual compositions will require higher-quality spectroscopy.

As discussed in Section \ref{sec:abun}, the absence of hydrogen is expected for a surviving D$^6$ WD, as it would have been transferred stably to the primary WD and ejected from the system prior to the explosion.  The non-detection of helium is unconstraining due to the relatively low surface temperatures.  In fact, in their modeling of GD~492, a hypervelocity star spectroscopically similar to our three candidates, \cite{radd18a} find significant helium is required even though it is not directly observable.

\begin{figure}
  \centering
  \includegraphics[width=\columnwidth]{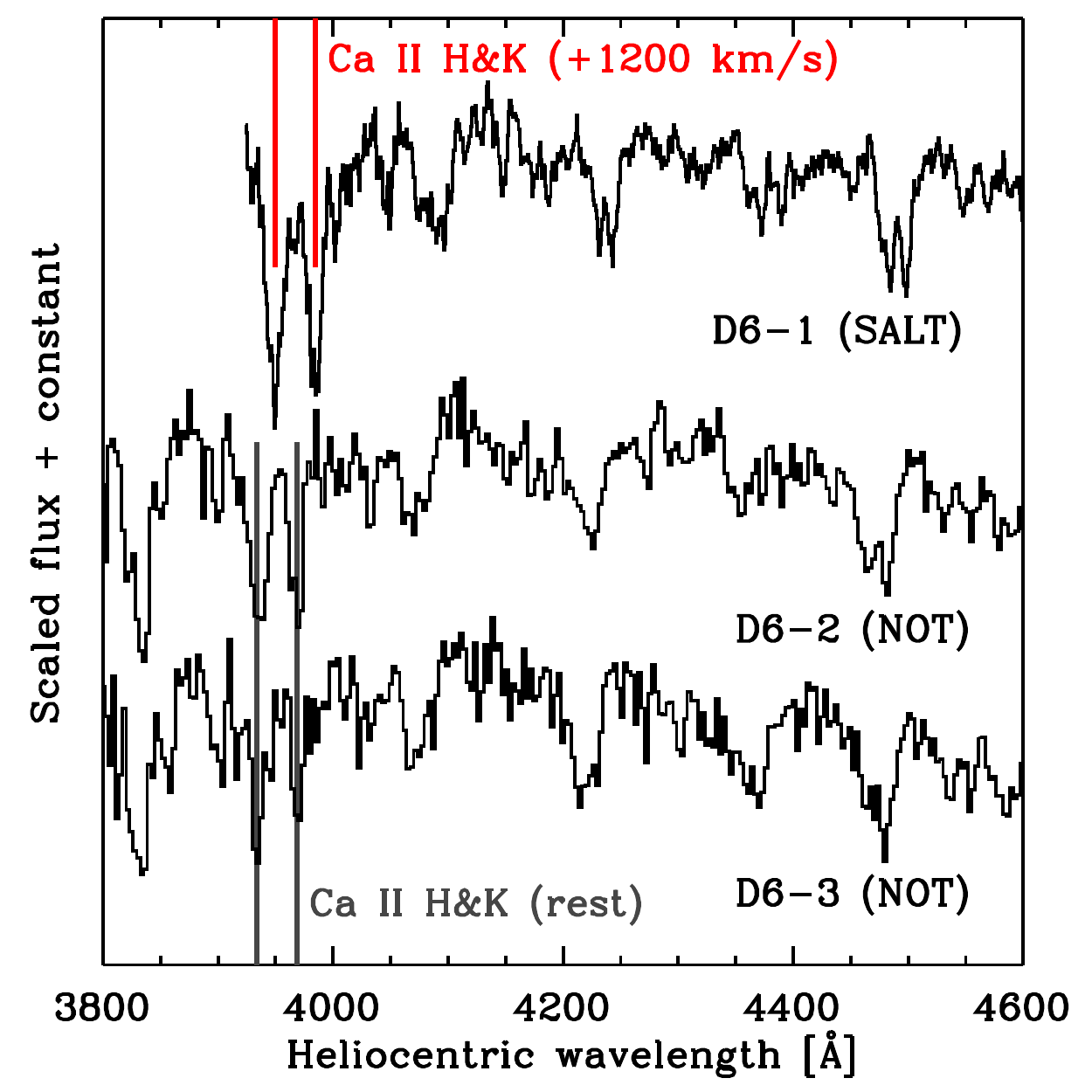}
  \caption{Zoom-in of the spectra for D6-1, D6-2, and D6-3.  Observed wavelengths have been transformed to the heliocentric frame but are otherwise uncorrected for RV shifts.  The Ca {\sc ii} H\&K lines for D6-1 are clearly shifted by $\unit[1200]{km \, s^{-1}}$, as shown by the red lines, while the same features are consistent with their rest wavelength values for D6-2 and D6-3.}
  \label{fig:zoomspecWD}
\end{figure}

Figure \ref{fig:zoomspecWD} shows a zoomed-in portion of the D6-1, D6-2, and D6-3 observed wavelength spectra, highlighting the region near the Ca {\sc ii} H\&K lines.  It is clear that D6-1 has a radial velocity shift of $\unit[1200]{km \, s^{-1}}$, as shown by the red lines.  However, D6-2 and D6-3 have RVs consistent with being $< \unit[100]{km \, s^{-1}}$.  These very low RVs cast doubt on the interpretation of these stars as hypervelocity stars: it seems unlikely that these stars would have a combination of very high proper motion velocities but very low RVs.  As a check, all the \emph{Gaia} proper motions were confirmed through examination of the fields of these candidates in epochs 1 and 2 of the Digitized Sky Survey, and also the Sloan Digital Sky Survey \citep{alam15a} for D6-2. In all cases the long baseline proper motions are consistent with \emph{Gaia}'s values.  If we assume that the measured parallaxes of D6-2 and D6-3 are instead systematically incorrect and that the transverse velocities of D6-2 and D6-3 are a more typical $\unit[100]{km \, s^{-1}}$, then the implied distances are $\sim \unit[100]{pc}$. This would suggest absolute magnitudes of $\emph{G}=12-13$.  Thus, a possible interpretation is that these two objects are faint, nearby white dwarfs and that the parallax measurements have extremely large systematic uncertainties.

On the other hand, the \emph{Gaia} noise values for D6-2 and D6-3 imply clean measurements.  Moreover, D6-1's extremely high RV makes its proper motion velocity credible.\footnote{D6-1's RV was confirmed on a subsequent night, ruling out the possibility that the high RV is due to orbital motion in a tight binary system.}  Finally, the fact that all three candidates, selected for their extreme proper motion velocities, are similar to each other and to GD~492, another hypervelocity star \citep{venn17a,radd18a,radd18b}, suggests that the proper motion velocities of D6-1, D6-2, and D6-3 are indeed very high.


\subsection{Posterior velocity distributions and orbital solutions}

\begin{figure}
  \centering
  \includegraphics[width=\columnwidth]{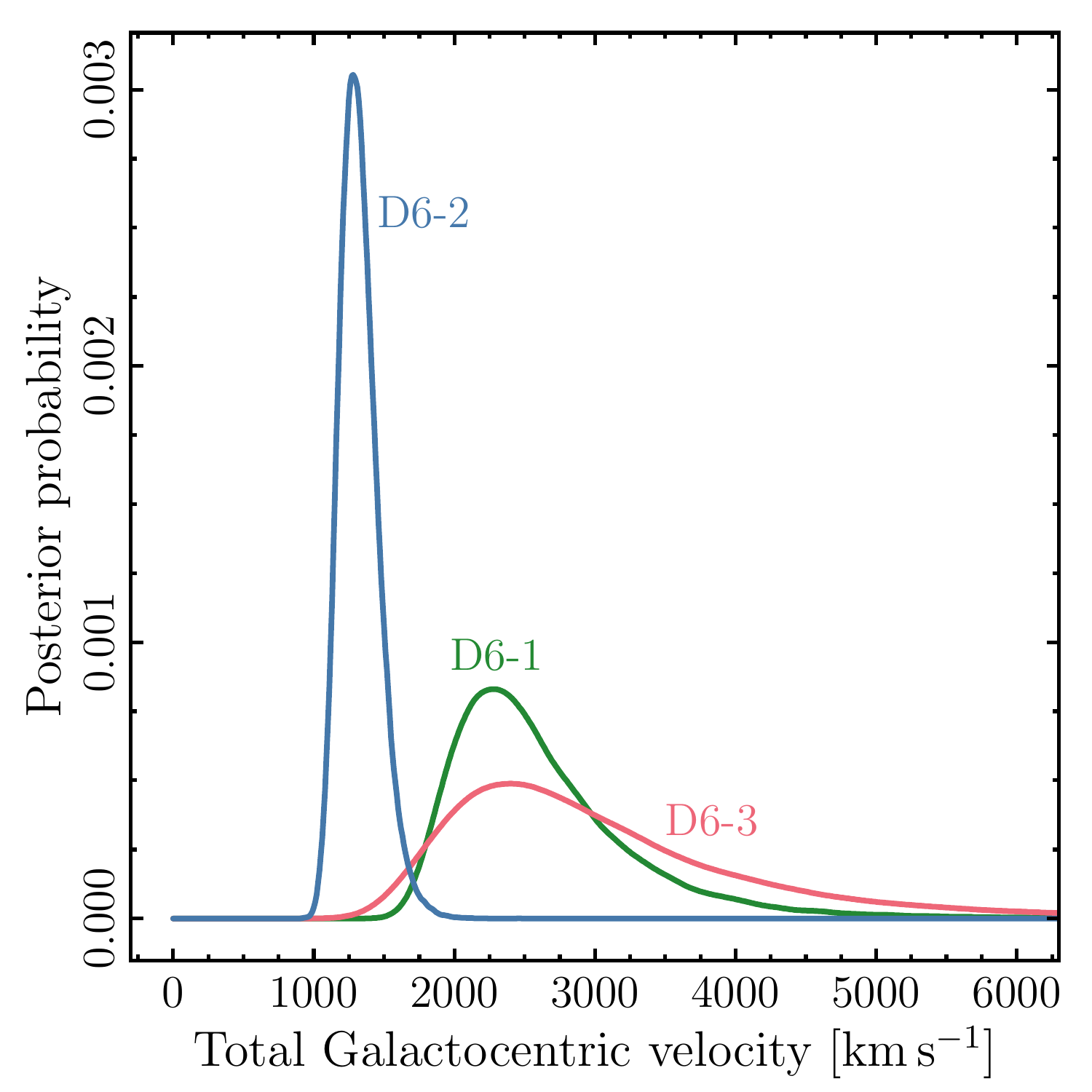}
  \caption{Posterior probabilities of total Galactocentric velocities for D6-1, D6-2, and D6-3.  An exponentially decreasing space density distance prior is used, and the parallax, proper motion, and RV errors are assumed to be normally distributed.}
  \label{fig:vpost}
\end{figure}

\begin{deluxetable}{c|ccc}
\tablecaption{Velocity information of the hypervelocity candidates \label{tab:vel}}
\tablehead{
\colhead{Nickname} &  \colhead{RV$_{\rm helio}$\tablenotemark{a}} & \colhead{$v_{\rm proper,helio}$\tablenotemark{b}} & \colhead{$v_{\rm Galacto}$\tablenotemark{c}} \\
\colhead{} & \colhead{($\unit[]{km \, s^{-1}}$)} & \colhead{($\unit[]{km \, s^{-1}}$)} & \colhead{($\unit[]{km \, s^{-1}}$)}   }
\startdata
D6-1 & $1200 \pm 40$ &$2200$ [$1400-6800$] &  $2300$ [$1600-6600$] \\
D6-2 & $20 \pm 60$ &$1200$ [$700-1500$] &  $1300$ [$1000-1900$] \\
D6-3 & $-20 \pm 80$ &$2400$ [$1700-11100$] &  $2400$ [$1400-9000$] \\
\enddata
\tablenotetext{a}{Heliocentric radial velocity, with 68.3\% uncertainties.}
\tablenotetext{b}{Heliocentric proper motion velocity, posterior maximum with 99.7\% credible intervals.}
\tablenotetext{c}{Total Galactocentric velocity, posterior maximum with 99.7\% credible intervals.}
\end{deluxetable}

To emphasize the extreme nature of D6-1, D6-2, and D6-3, Figure \ref{fig:vpost} shows the posterior distributions of their total Galactocentric velocities.  These posteriors were derived by sampling the distance posteriors (equation \ref{eqn:post}), as well as sampling the RVs and proper motions within their assumed Gaussian uncertainties, and applying a heliocentric to Galactocentric coordinate transformation with \texttt{astropy} \citep{astr18a}.

Further velocity information is listed in Table \ref{tab:vel}.  D6-1, D6-2, and D6-3 have high probabilities of being three of the fastest known stars in the Milky Way, possibly only surpassed by pulsars kicked from core collapse SNe and the stars in close orbit around the Galactic center.

\begin{figure*}
  \centering
  \includegraphics[width=\textwidth]{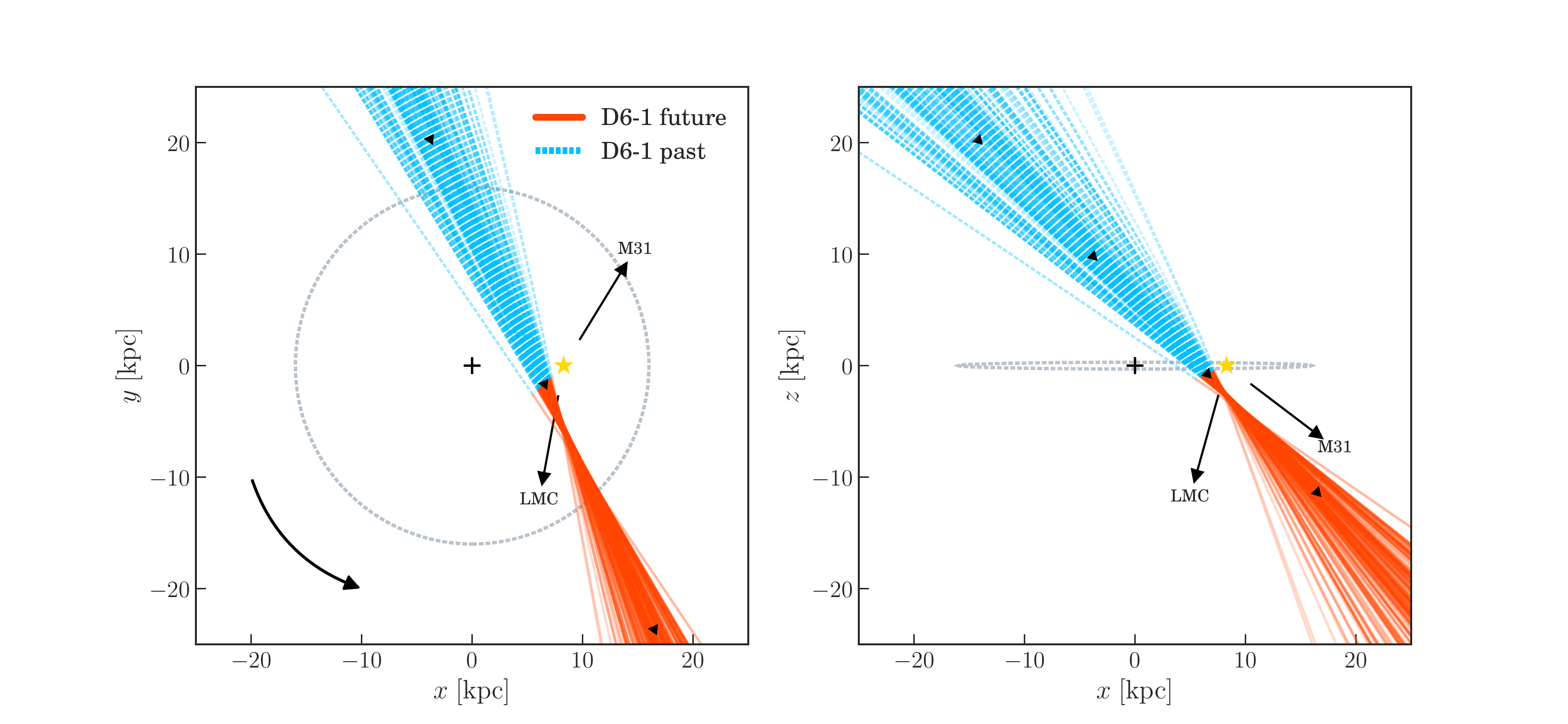}
  \caption{Sample realizations of D6-1's orbital solutions (past trajectories in blue; future trajectories in red).  The Milky Way's thin disk is overlaid in gray contours.  A face-on view is shown in the left panel; the right panel shows an edge-on perspective.  The Sun's location is marked with a star, and the Galaxy's rotation and the directions to the LMC and M31 are denoted with arrows.}
  \label{fig:orbD6-1}
\end{figure*}

\begin{figure*}
  \centering
  \includegraphics[width=\textwidth]{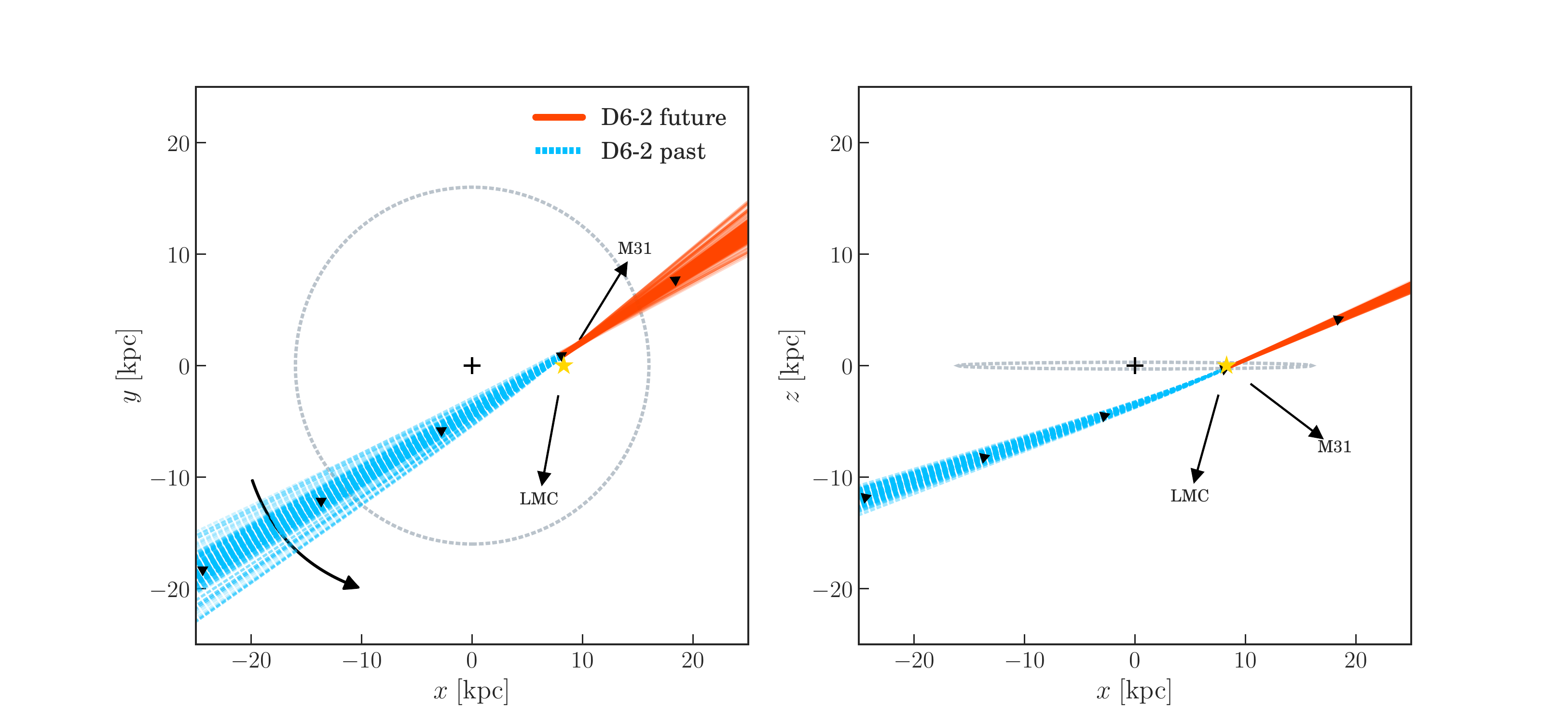}
  \caption{Same as Figure \ref{fig:orbD6-1}, but for D6-2's orbital solutions.}
  \label{fig:orbD6-2}
\end{figure*}

\begin{figure*}
  \centering
  \includegraphics[width=\textwidth]{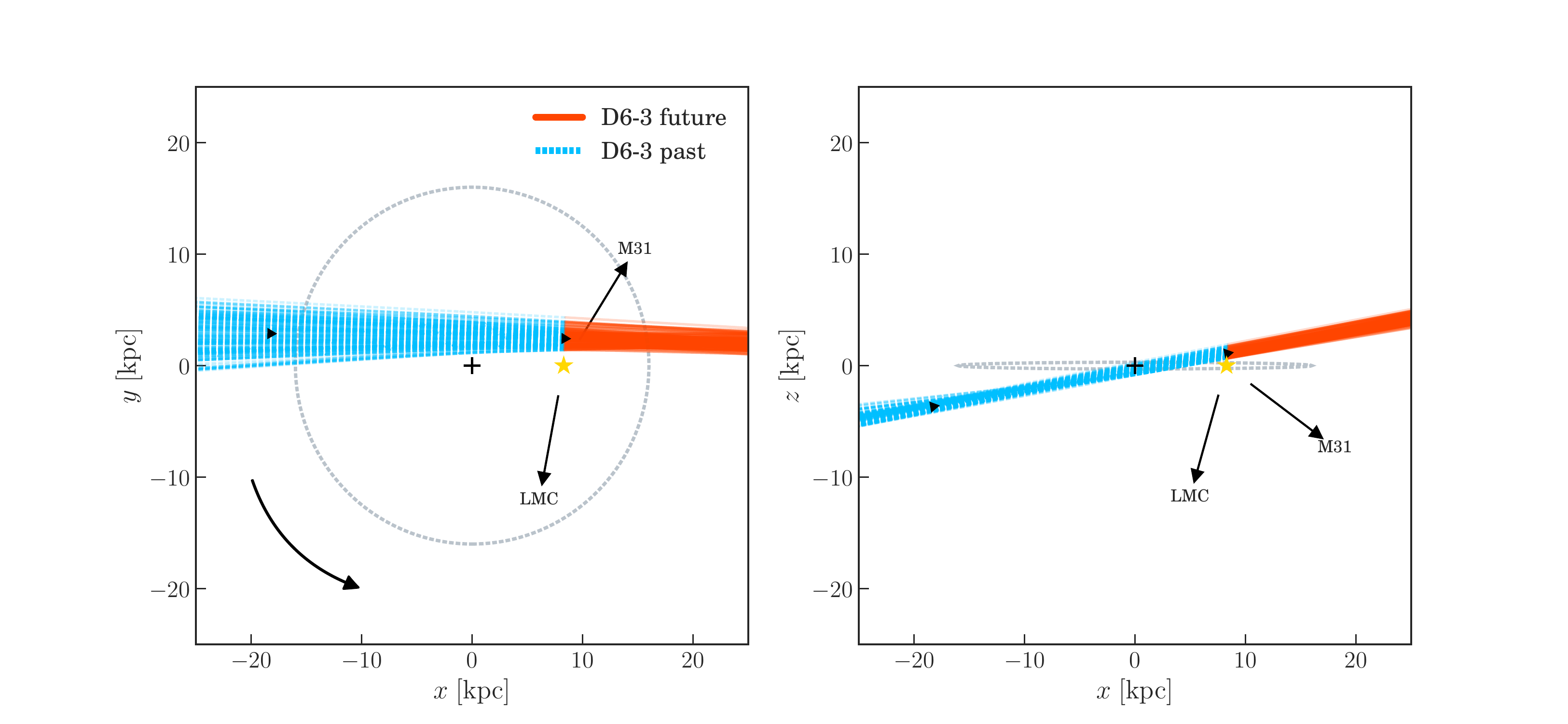}
  \caption{Same as Figure \ref{fig:orbD6-1}, but for D6-3's orbital solutions.}
  \label{fig:orbD6-3}
\end{figure*}

Samples of the orbital solutions for the candidate hypervelocity stars are shown in Figures \ref{fig:orbD6-1}, \ref{fig:orbD6-2}, and \ref{fig:orbD6-3}, calculated with \texttt{galpy} \citep{bovy15a} and assuming an MWPotential2014 gravitational potential and an exponentially decreasing space density distance prior as before.   It is clear that all three candidates are unbound from the Milky Way and that almost none of the orbital solutions passes near the Galactic center.  Taken as a group, it is highly unlikely that the hypervelocity nature of these stars is due to Galactic center ejection.


\subsection{Color-magnitude diagram and possible interpretation}
\label{sec:cmd}

\begin{figure}
  \centering
  \includegraphics[width=\columnwidth]{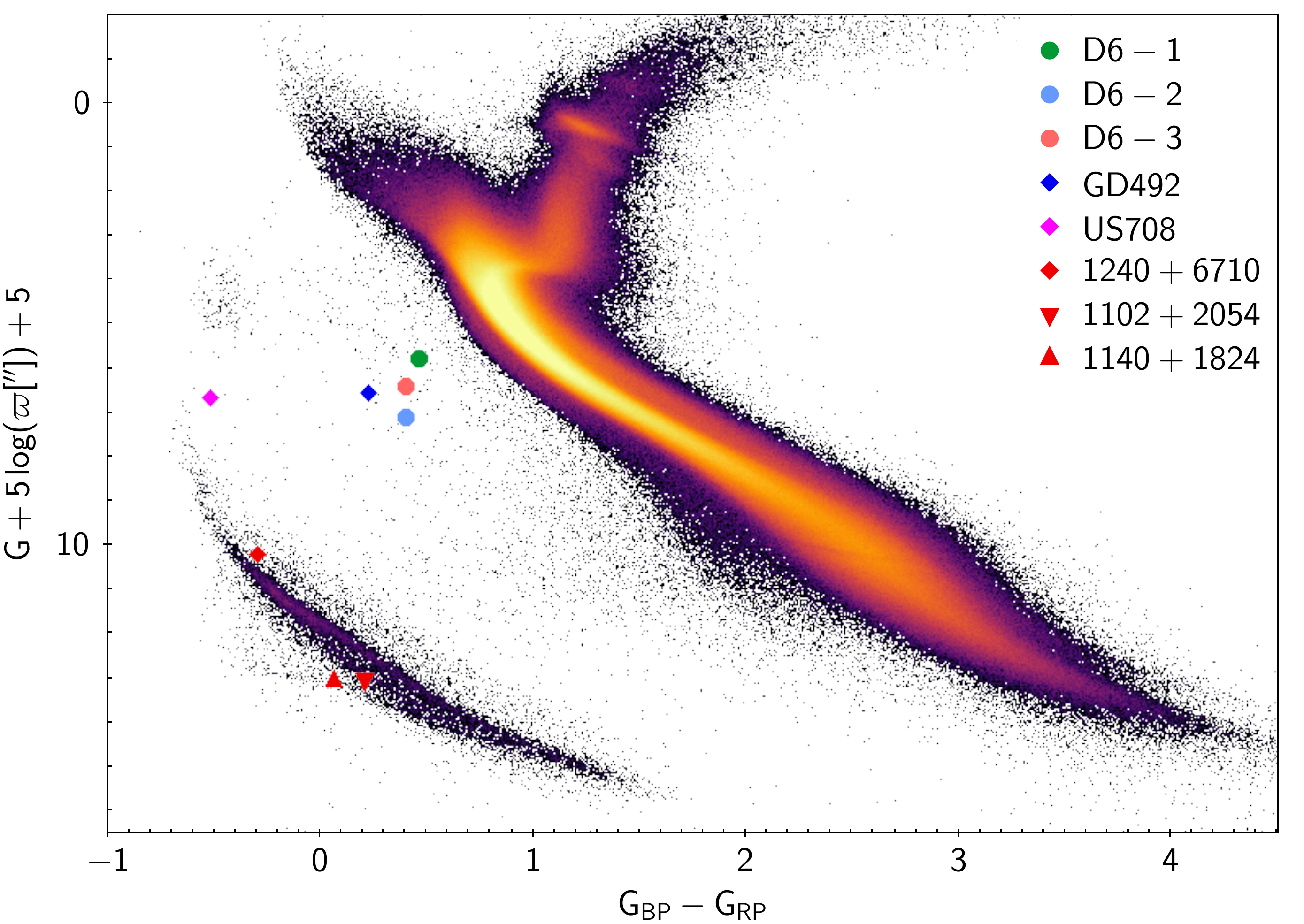}
  \caption{Color-magnitude diagram of the three hypervelocity candidates, GD~492, US~708, and three chemically peculiar WDs (colored symbols).  Black circles and colored regions show reliably measured stars from \emph{Gaia}.}
  \label{fig:cmd}
\end{figure}

In Figure \ref{fig:cmd}, we show a color-magnitude diagram using \emph{Gaia}'s $G_{\rm BP}-G_{\rm RP}$ color and the absolute \emph{Gaia} $G$-band magnitude, assuming the measured values of the parallax.  Black points and colored regions show a random sub-sample of $5.7\E{6}$ stars from \emph{Gaia} DR2 with accurate parallaxes ($\varpi/\sigma_\varpi>30$), clean astrometry (\texttt{astrometric\_excess\_noise} $<1$) and low reddening ($|b|>30^\circ$).  Our three hypervelocity candidates, GD~492 \citep{venn17a}, US~708 \citep{hirs05a,just09a,geie15a}, and the three peculiar SDSS WDs from Figure \ref{fig:specWD} are shown as colored symbols.

D6-1, D6-2, and D6-3 form a relatively tight grouping in color-magnitude space that includes GD~492.  These four stars currently have radii between typical WDs and main sequence stars, similar to subdwarf stars; GD~492's radius is estimated to be $0.2 \, R_\odot$ \citep{radd18b}.  However, if they had previously been helium-rich subdwarf companions of exploding WDs in SN~Ia systems, Figure \ref{fig:mvsv} shows that their velocities would be well below $\unit[1000]{km \, s^{-1}}$.  There have been some suggestions that GD~492 is instead the bound WD remnant of a SN~Iax explosion that has received a large kick due to asymmetric mass ejection or due to the disruption of the binary system upon instantaneous mass loss \citep{venn17a,radd18a,radd18b}.  However, the predicted kicks in these systems due to asymmetric mass ejection range from only ten to several hundred $\unit[]{km \, s^{-1}}$ \citep{fink14a,long14a}, far below the observed velocities of the D$^6$ stars.  The orbital velocity of a Chandrasekhar-mass WD in a tight binary with a helium star can approach GD~492's velocity, but it cannot explain the even higher velocities of D6-1, D6-2, and D6-3.

These three stars, and possibly also GD~492, may instead be the surviving companion WDs of D$^6$ SNe~Ia.  While they are clearly not typical WDs now, mechanisms discussed in Section \ref{sec:lum} that were active during the phase of dynamical mass transfer prior to the SN~Ia and the post-explosion evolution may have deposited enough energy to lift the degeneracy of the outer layers and cause the WDs to temporarily appear as subdwarf stars just after the explosion.

Our expectation was that most of the excess energy would be deposited near the surface of the WD, where it would be radiated away on relatively short timescales, so that the stars would quickly return to being dim, typical-looking WDs.  This could indeed be true, making it difficult to observe the majority of the runaway WDs in the Solar neighborhood.  However, a small fraction of these stars will have experienced SNe~Ia much more recently than the average runaway WD, possibly rendering them still bright enough to observe.

A similar calculation to the one described in Section \ref{sec:numest} yields $300-400$ runaway WDs within the $\sim \unit[65]{kpc^3}$ volume in which we observed D6-1, D6-2, and D6-3.  If we include GD~492, we have observed 1\% of the potential nearby runaway WDs, which corresponds to stars ejected more recently than $ \sim \unit[4\E{4}]{yr}$ (Fig.\ \ref{fig:agecdf}).  Thus, D6-1, D6-2, D6-3, and possibly GD~492 may just represent the small portion of runaway WDs that have been violently altered by SNe~Ia so recently they have yet to evolve back into typical-looking WDs.

US~708, a hypervelocity helium-rich subdwarf, sits blueward of this group.  It is possible that, as stars like D6-1, D6-2, D6-3, and GD~492 radiate the deposited energy and contract, the unseen helium that \cite{radd18a} require in their best-fit models of GD~492's spectrum becomes directly observable as the photosphere becomes hotter.  Thus, these stars could evolve to appear like US~708 on their way back to the WD cooling track.  Future observations and detailed stellar evolution calculations and spectral modeling will help to test all of these intriguing possibilities.  


\subsection{D6-2's association with SN remnant G70.0-21.5}

Motivated by the possibility that these four stars may have been ejected from the sites of SNe~Ia $\sim \unit[4\E{4}]{yr}$ ago, we search for existing SN remnants along their past orbital solutions up to their positions $\sim\unit[10^5]{yr}$ ago.  We use the online catalogs of D.A.\ Green\footnote{\url{https://www.mrao.cam.ac.uk/surveys/snrs}} and G.\ Ferrand\footnote{\url{http://www.physics.umanitoba.ca/snr/SNRcat}}, as well as the Open Supernova Catalog\footnote{\url{https://sne.space}} \citep{guil17a}. Additionally, we searched data from the {\it ROSAT} All-Sky Survey for SN remnant sources that emit soft X-rays.

For stars D6-1, D6-3, and GD~492 we find no SN remnant candidate sources along the past velocity vectors. We note that this does not rule out their association with a SN, because SN remnants can dissipate into the interstellar medium on timescales ranging from a few thousand years up to a few hundred thousand years, depending on the remnant.  Furthermore, the regions around these three stars lack high-quality H$\alpha$ imaging, so limits on the non-existence of remnants are not constraining.

\begin{figure*}
  \centering
  \includegraphics[width=\textwidth]{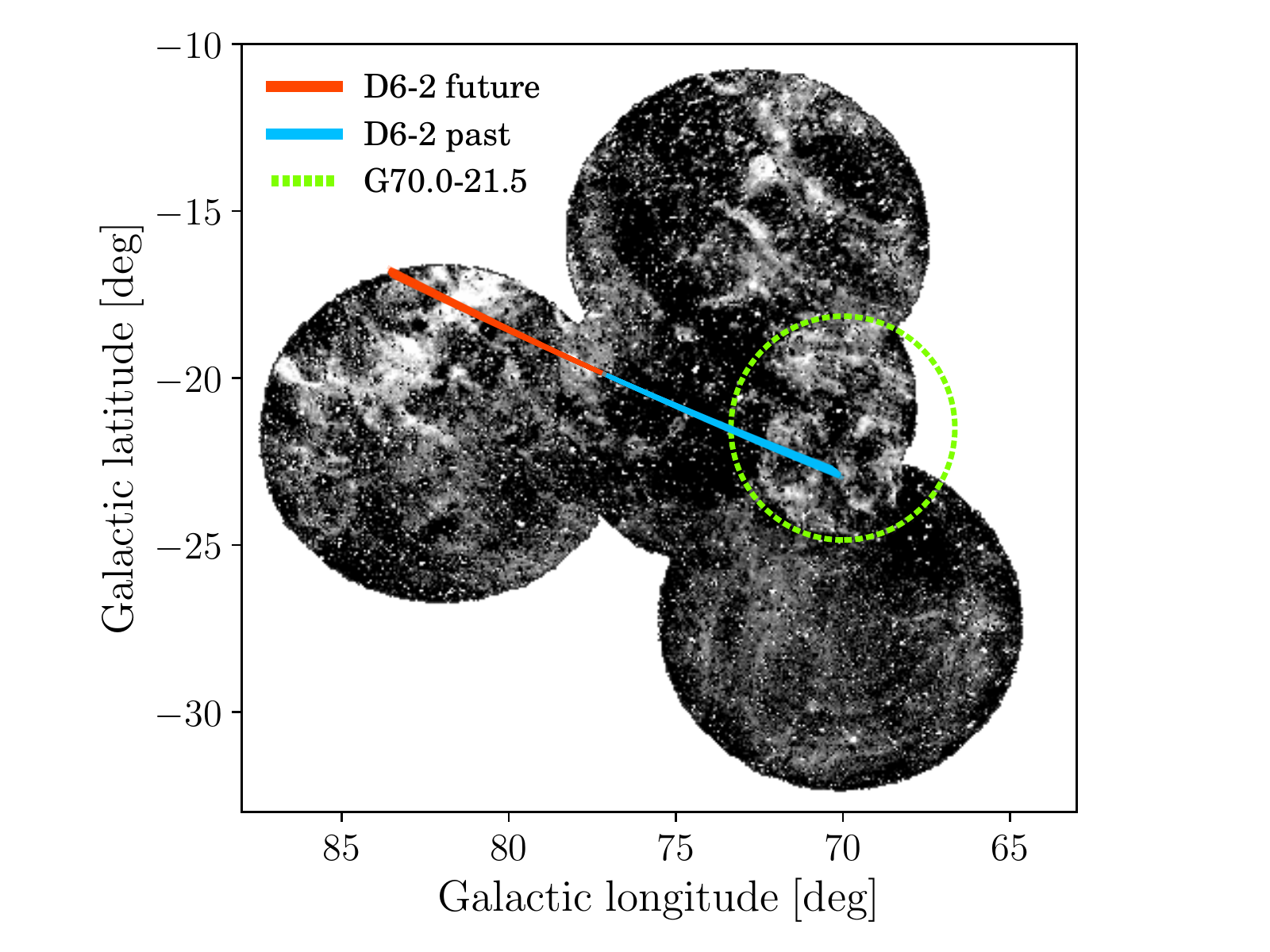}
  \caption{The orbital solution of D6-2 overlaid on H$\alpha$ images from Virginia Tech Spectral Line Survey (VTSS; \citealt{denn98a}).  The blue and red trajectories extend $\unit[9\E{4}]{yr}$ into D6-2's past and future, respectively.  The green circle encompasses the remnant of G70.0-21.5.}
  \label{fig:g70}
\end{figure*}

However, as shown in Figure \ref{fig:g70}, D6-2's position $\unit[9\E{4}]{yr}$ in the past places it \unit[0.9]{deg} from the approximate geometric center of the faint, old SN remnant G70.0-21.5.  This remnant, first identified by \citet{fese15a}, consists of a shell of H$\alpha$ filaments, along with several other spectral lines only associated with slowly moving, radiative shocks from an old SN remnant. Additionally, \citet{fese15a} report faint {\it ROSAT} X-ray emission from near the center, another  indication of a SN remnant.

Of the known remnants, G70.0-21.5 lies furthest from the Galactic plane, suggesting it was produced by a SN~Ia.  The distance to G70.0-21.5 inferred from its shock velocities is $\unit[1-2]{kpc}$, consistent with D6-2's parallax-measured distance of $ \unit[1.0 \pm 0.1 ]{kpc} $.  \cite{fese15a} also conclude that the remnant is quite old, perhaps between several $10^4$ and $\unit[10^5]{yr}$.  The probability of a chance alignment between D6-2's past position and the projected center of an unassociated SN remnant that is at a consistent distance and has a consistent age is likely very small, especially given the lack of any other obvious SN remnants in the VTSS images of this region.  However, this probability should be quantified in future work.

At a distance of $\unit[1]{kpc}$, G70.0-21.5's height below the Galactic plane is $\unit[400]{pc}$, matching D6-2's height $\unit[10^5]{yr}$ ago.  D6-1 and D6-3 were farther from the plane at that time: $700$ and $\unit[1200]{pc}$, respectively.  Such offsets are reasonable since SNe~Ia can occur in old stellar populations and the thick disk's scale height is $\unit[900]{pc}$ \citep{blan16a}.  Furthermore, D$^6$ survivors that were recently ejected from closer to the Galactic plane would suffer higher extinction and could be unobservable.  Thus, these three D$^6$ candidates may represent more than just 1\% of the $300-400$ expected survivors within $\unit[2.5]{kpc}$, after accounting for those that remain obscured at low Galactic latitudes.

While D6-2 did not pass exactly through G70.0-21.5's center as reported by \cite{fese15a}, the approximate determination of the center of this aspherical remnant makes such a comparison difficult.  Moreover, G70.0-21.5's advanced age means that its appearance is heavily influenced by the structure of the inhomogeneous surrounding interstellar medium, further complicating determination of the site of the explosion.  We thus do not regard the $\unit[0.9]{deg}$ offset between D6-2's past position and \cite{fese15a}'s reported center as evidence against the association.

It is not possible to rule out other SN remnants present in this field that are even older, fainter, and more diffuse than G70.0-21.5.  At some point, as SN remnants age, they simply dissolve away into the interstellar medium, below any possible threshold of detection. Nonetheless, the close association between D6-2 and a known SN remnant, G70.0-21.5, is quite tantalizing. Though the remnant is at a very advanced evolutionary stage, it may be possible to follow up the X-ray emission with deeper observations to determine the ejecta abundances, and thus, the SN type.  We note that while none of the other hypervelocity runaway candidates are associated with known SN remnants, they also all lack high-quality H$\alpha$ imaging, which is how G70.0-21.5 was discovered.  Future observations will ascertain if any of the remaining candidates' remnants are observable but are so faint they have been missed by previous lower-quality searches.


\section{Conclusions}
\label{sec:conc}

We have searched \emph{Gaia}'s second data release for hypervelocity runaway WDs that survived dynamically driven double-degenerate double-detonation (D$^6$) SNe~Ia.  We followed up seven candidates with ground-based telescopes.  Of these, three are consistent with being hypervelocity runaway stars that were previously the WD companions to primary WDs that exploded as SNe~Ia.  One candidate is also closely associated with a faint, old SN remnant at a distance consistent with the candidate's measured parallax.

One lingering puzzle is the very low RV measured for two of the hypervelocity stars.  However, given these two stars' close photometric and spectroscopic match to the third star, which does have a large RV, and the association of one of the low RV stars with a SN remnant, the peculiar RVs may just be a statistical fluctuation due to small numbers.  Future detection and characterization of more D$^6$ survivors will hopefully ease this tension.

While the candidates are much brighter and have larger radii than expected, plausible mechanisms exist that may have changed the appearance of these hypervelocity runaway stars for a short time following the SN~Ia explosion.  If validated by future high-resolution spectroscopy and detailed stellar evolution calculations and atmospheric modeling, these stars would confirm the success of the D$^6$ scenario in producing SNe~Ia.  Future study of these candidate D$^6$ survivors would also shed important insight on the physics of double WD binaries and the aftermath of SN~Ia explosions, including tidal heating, stellar ablation, and ejecta deposition.

The increased luminosity of the three D$^6$ candidates raises the possibility that such survivors could be observable within relatively close and young SN remnants such as Tycho, Kepler, and SN 1006.  However, previous searches within these remnants, including one designed to look for hot surviving WD companions, and our search using \emph{Gaia} DR2 have failed to find any convincing candidates.  It is possible that D$^6$ survivors do exist within these remnants but are presently too faint or too red or blue to be easily disentangled from the other stars.  Future analysis and modeling will help to discover or constrain the existence of surviving WDs within these remnants.

Several candidates, undiscussed here, remain to be followed up.  Furthermore, given the large parallax uncertainties near the magnitude limit of DR2, it is likely that more such stars exist but are hiding in the data.  Dedicated work and future \emph{Gaia} data releases may be able to tease out more candidates in the coming years.


\acknowledgments

We thank Josh Bloom, Brian Metzger, Alison Miller, Peter Nugent, and Eliot Quataert for discussions, and the referee for helpful comments.  K.J.S.\ is supported by NASA through the Astrophysics Theory Program (NNX17AG28G).  D.B.\ thanks the UK Science and Technology Facilities Council for supporting his PhD.  We thank Encarni Romero Colmenero and Petri Vaisanen for the Director's Discretionary Time award on SALT. S.W.J.\ is supported in part by NSF award AST-1615455.  The UCSC group is supported in part by NASA grant NNG17PX03C, NSF grant AST-1518052, the Gordon \& Betty Moore Foundation, the Heising-Simons Foundation, and by fellowships from the Alfred P.\ Sloan Foundation and the David and Lucile Packard Foundation to R.J.F.  M.F.\ is supported by a Royal Society - Science Foundation Ireland University Research Fellowship.  K.M.\ is supported by STFC through an Ernest Rutherford fellowship.  W.E.K.\ is supported by an ESO Fellowship and the Excellence Cluster Universe, Technische Universit{\"a}t
M{\"u}nchen, Boltzmannstrasse 2, D-85748 Garching, Germany.  Support for J.S.\ is provided by NASA through Hubble Fellowship grant \#HST-HF2-51382.001-A awarded by the Space Telescope Science Institute, which is operated by the Association of Universities for Research in Astronomy, Inc., for NASA, under contract NAS5-26555.  We acknowledge the ``Observational Signatures of Type Ia Supernova Progenitors III'' workshop at the Lorentz Center, where some of the preparatory work for this study was performed.

This work has made use of data from the European Space Agency (ESA) mission {\it Gaia} (\url{https://www.cosmos.esa.int/gaia}), processed by the {\it Gaia} Data Processing and Analysis Consortium (DPAC,
\url{https://www.cosmos.esa.int/web/gaia/dpac/consortium}). Funding for the DPAC has been provided by national institutions, in particular the institutions participating in the {\it Gaia} Multilateral Agreement.

The NOT data was obtained through NUTS, which is supported in part by the Instrument Center for Danish Astrophysics (IDA).  Support also comes from the European Research Council under the European Union's Seventh Framework Programme (FP/2007-2013) / ERC Grant Agreement n. 320964 (WDTracer).  Some of the observations reported in this paper were obtained with the Southern African Large Telescope (SALT).  PyRAF is a product of the Space Telescope Science Institute, which is operated by AURA for NASA.

The Pan-STARRS1 Surveys (PS1) and the PS1 public science archive have been made possible through contributions by the Institute for Astronomy, the University of Hawaii, the Pan-STARRS Project Office, the Max-Planck Society and its participating institutes, the Max Planck Institute for Astronomy, Heidelberg and the Max Planck Institute for Extraterrestrial Physics, Garching, The Johns Hopkins University, Durham University, the University of Edinburgh, the Queen's University Belfast, the Harvard-Smithsonian Center for Astrophysics, the Las Cumbres Observatory Global Telescope Network Incorporated, the National Central University of Taiwan, the Space Telescope Science Institute, the National Aeronautics and Space Administration under Grant No.\ NNX08AR22G issued through the Planetary Science Division of the NASA Science Mission Directorate, the National Science Foundation Grant No.\ AST-1238877, the University of Maryland, Eotvos Lorand University (ELTE), the Los Alamos National Laboratory, and the Gordon and Betty Moore Foundation.

The national facility capability for SkyMapper has been funded through ARC LIEF grant LE130100104 from the Australian Research Council, awarded to the University of Sydney, the Australian National University, Swinburne University of Technology, the University of Queensland, the University of Western Australia, the University of Melbourne, Curtin University of Technology, Monash University and the Australian Astronomical Observatory. SkyMapper is owned and operated by The Australian National University's Research School of Astronomy and Astrophysics. The survey data were processed and provided by the SkyMapper Team at ANU. The SkyMapper node of the All-Sky Virtual Observatory (ASVO) is hosted at the National Computational Infrastructure (NCI). Development and support the SkyMapper node of the ASVO has been funded in part by Astronomy Australia Limited (AAL) and the Australian Government through the Commonwealth's Education Investment Fund (EIF) and National Collaborative Research Infrastructure Strategy (NCRIS), particularly the National eResearch Collaboration Tools and Resources (NeCTAR) and the Australian National Data Service Projects (ANDS).


\software{\texttt{rvsao} \citep{kurt98a}, \texttt{astropy} \citep{astr18a}, \texttt{galpy} \citep{bovy15a}, \texttt{MESA} (v10108; \citealt{paxt11,paxt13,paxt15a,paxt18a}), \texttt{PySALT} (\url{http://pysalt.salt.ac.za}; \citealt{craw10a}), \texttt{PyRAF} (\url{http://www.stsci.edu/institute/software_hardware/pyraf}), \texttt{ALFOSCGUI} (\url{http://sngroup.oapd.inaf.it/foscgui.html})}




\begin{thebibliography}{}
\expandafter\ifx\csname natexlab\endcsname\relax\def\natexlab#1{#1}\fi
\providecommand{\url}[1]{\href{#1}{#1}}
\providecommand{\dodoi}[1]{doi:~\href{http://doi.org/#1}{\nolinkurl{#1}}}
\providecommand{\doeprint}[1]{\href{http://ascl.net/#1}{\nolinkurl{http://ascl.net/#1}}}
\providecommand{\doarXiv}[1]{\href{https://arxiv.org/abs/#1}{\nolinkurl{https://arxiv.org/abs/#1}}}

\bibitem[{{Alam} {et~al.}(2015){Alam}, {Albareti}, {Allende Prieto}, {Anders},
  {Anderson}, {Anderton}, {Andrews}, {Armengaud}, {Aubourg}, {Bailey}, \&
  et~al.}]{alam15a}
{Alam}, S., {Albareti}, F.~D., {Allende Prieto}, C., {et~al.} 2015, \apjs, 219,
  12, \dodoi{10.1088/0067-0049/219/1/12}

\bibitem[{{Astraatmadja} \& {Bailer-Jones}(2016)}]{astr16a}
{Astraatmadja}, T.~L., \& {Bailer-Jones}, C.~A.~L. 2016, \apj, 832, 137,
  \dodoi{10.3847/0004-637X/832/2/137}

\bibitem[{{Bergeron} {et~al.}(2011){Bergeron}, {Wesemael}, {Dufour},
  {Beauchamp}, {Hunter}, {Saffer}, {Gianninas}, {Ruiz}, {Limoges}, {Dufour},
  {Fontaine}, \& {Liebert}}]{berg11a}
{Bergeron}, P., {Wesemael}, F., {Dufour}, P., {et~al.} 2011, \apj, 737, 28,
  \dodoi{10.1088/0004-637X/737/1/28}

\bibitem[{{Bildsten} {et~al.}(2007){Bildsten}, {Shen}, {Weinberg}, \&
  {Nelemans}}]{bild07}
{Bildsten}, L., {Shen}, K.~J., {Weinberg}, N.~N., \& {Nelemans}, G. 2007,
  \apjl, 662, L95, \dodoi{10.1086/519489}

\bibitem[{{Bland-Hawthorn} \& {Gerhard}(2016)}]{blan16a}
{Bland-Hawthorn}, J., \& {Gerhard}, O. 2016, \araa, 54, 529,
  \dodoi{10.1146/annurev-astro-081915-023441}

\bibitem[{{Blondin} {et~al.}(2017){Blondin}, {Dessart}, {Hillier}, \&
  {Khokhlov}}]{blon17a}
{Blondin}, S., {Dessart}, L., {Hillier}, D.~J., \& {Khokhlov}, A.~M. 2017,
  \mnras, 470, 157, \dodoi{10.1093/mnras/stw2492}

\bibitem[{{Bloom} {et~al.}(2012){Bloom}, {Kasen}, {Shen}, {Nugent}, {Butler},
  {Graham}, {Howell}, {Kolb}, {Holmes}, {Haswell}, {Burwitz}, {Rodriguez}, \&
  {Sullivan}}]{bloo12}
{Bloom}, J.~S., {Kasen}, D., {Shen}, K.~J., {et~al.} 2012, \apjl, 744, L17,
  \dodoi{10.1088/2041-8205/744/2/L17}

\bibitem[{{Bovy}(2015)}]{bovy15a}
{Bovy}, J. 2015, \apjs, 216, 29, \dodoi{10.1088/0067-0049/216/2/29}

\bibitem[{{Brown}(2015)}]{brow15a}
{Brown}, W.~R. 2015, \araa, 53, 15, \dodoi{10.1146/annurev-astro-082214-122230}

\bibitem[{{Brown} {et~al.}(2016){Brown}, {Kilic}, {Kenyon}, \&
  {Gianninas}}]{brow16b}
{Brown}, W.~R., {Kilic}, M., {Kenyon}, S.~J., \& {Gianninas}, A. 2016, \apj,
  824, 46, \dodoi{10.3847/0004-637X/824/1/46}

\bibitem[{{Bulla} {et~al.}(2016){Bulla}, {Sim}, {Pakmor}, {Kromer},
  {Taubenberger}, {R{\"o}pke}, {Hillebrandt}, \& {Seitenzahl}}]{bull16a}
{Bulla}, M., {Sim}, S.~A., {Pakmor}, R., {et~al.} 2016, \mnras, 455, 1060,
  \dodoi{10.1093/mnras/stv2402}

\bibitem[{{Burkart} {et~al.}(2013){Burkart}, {Quataert}, {Arras}, \&
  {Weinberg}}]{burk13a}
{Burkart}, J., {Quataert}, E., {Arras}, P., \& {Weinberg}, N.~N. 2013, \mnras,
  433, 332, \dodoi{10.1093/mnras/stt726}

\bibitem[{{Chambers} {et~al.}(2016){Chambers}, {Magnier}, {Metcalfe},
  {Flewelling}, {Huber}, {Waters}, {Denneau}, {Draper}, {Farrow}, {Finkbeiner},
  {Holmberg}, {Koppenhoefer}, {Price}, {Saglia}, {Schlafly}, {Smartt},
  {Sweeney}, {Wainscoat}, {Burgett}, {Grav}, {Heasley}, {Hodapp}, {Jedicke},
  {Kaiser}, {Kudritzki}, {Luppino}, {Lupton}, {Monet}, {Morgan}, {Onaka},
  {Stubbs}, {Tonry}, {Banados}, {Bell}, {Bender}, {Bernard}, {Botticella},
  {Casertano}, {Chastel}, {Chen}, {Chen}, {Cole}, {Deacon}, {Frenk},
  {Fitzsimmons}, {Gezari}, {Goessl}, {Goggia}, {Goldman}, {Grebel}, {Hambly},
  {Hasinger}, {Heavens}, {Heckman}, {Henderson}, {Henning}, {Holman}, {Hopp},
  {Ip}, {Isani}, {Keyes}, {Koekemoer}, {Kotak}, {Long}, {Lucey}, {Liu},
  {Martin}, {McLean}, {Morganson}, {Murphy}, {Nieto-Santisteban}, {Norberg},
  {Peacock}, {Pier}, {Postman}, {Primak}, {Rae}, {Rest}, {Riess}, {Riffeser},
  {Rix}, {Roser}, {Schilbach}, {Schultz}, {Scolnic}, {Szalay}, {Seitz},
  {Shiao}, {Small}, {Smith}, {Soderblom}, {Taylor}, {Thakar}, {Thiel},
  {Thilker}, {Urata}, {Valenti}, {Walter}, {Watters}, {Werner}, {White},
  {Wood-Vasey}, \& {Wyse}}]{cham16a}
{Chambers}, K.~C., {Magnier}, E.~A., {Metcalfe}, N., {et~al.} 2016,
  arXiv:1612.05560.
\newblock \doarXiv{1612.05560}

\bibitem[{{Chayer} {et~al.}(1995{\natexlab{a}}){Chayer}, {Fontaine}, \&
  {Wesemael}}]{chay95a}
{Chayer}, P., {Fontaine}, G., \& {Wesemael}, F. 1995{\natexlab{a}}, \apjs, 99,
  189, \dodoi{10.1086/192184}

\bibitem[{{Chayer} {et~al.}(1995{\natexlab{b}}){Chayer}, {Vennes}, {Pradhan},
  {Thejll}, {Beauchamp}, {Fontaine}, \& {Wesemael}}]{chay95b}
{Chayer}, P., {Vennes}, S., {Pradhan}, A.~K., {et~al.} 1995{\natexlab{b}},
  \apj, 454, 429, \dodoi{10.1086/176494}

\bibitem[{{Chiotellis} {et~al.}(2012){Chiotellis}, {Schure}, \&
  {Vink}}]{chio12a}
{Chiotellis}, A., {Schure}, K.~M., \& {Vink}, J. 2012, \aap, 537, A139,
  \dodoi{10.1051/0004-6361/201014754}

\bibitem[{{Colgate} \& {McKee}(1969)}]{cm69}
{Colgate}, S.~A., \& {McKee}, C. 1969, \apj, 157, 623, \dodoi{10.1086/150102}

\bibitem[{{Crawford} {et~al.}(2010){Crawford}, {Still}, {Schellart}, {Balona},
  {Buckley}, {Dugmore}, {Gulbis}, {Kniazev}, {Kotze}, {Loaring}, {Nordsieck},
  {Pickering}, {Potter}, {Romero Colmenero}, {Vaisanen}, {Williams}, \&
  {Zietsman}}]{craw10a}
{Crawford}, S.~M., {Still}, M., {Schellart}, P., {et~al.} 2010, in \procspie,
  Vol. 7737, Observatory Operations: Strategies, Processes, and Systems III,
  773725

\bibitem[{{Dan} {et~al.}(2015){Dan}, {Guillochon}, {Br{\"u}ggen},
  {Ramirez-Ruiz}, \& {Rosswog}}]{dan15a}
{Dan}, M., {Guillochon}, J., {Br{\"u}ggen}, M., {Ramirez-Ruiz}, E., \&
  {Rosswog}, S. 2015, \mnras, 454, 4411, \dodoi{10.1093/mnras/stv2289}

\bibitem[{{Dan} {et~al.}(2014){Dan}, {Rosswog}, {Br{\"u}ggen}, \&
  {Podsiadlowski}}]{dan14a}
{Dan}, M., {Rosswog}, S., {Br{\"u}ggen}, M., \& {Podsiadlowski}, P. 2014,
  \mnras, 438, 14, \dodoi{10.1093/mnras/stt1766}

\bibitem[{{Dan} {et~al.}(2011){Dan}, {Rosswog}, {Guillochon}, \&
  {Ramirez-Ruiz}}]{dan11}
{Dan}, M., {Rosswog}, S., {Guillochon}, J., \& {Ramirez-Ruiz}, E. 2011, \apj,
  737, 89, \dodoi{10.1088/0004-637X/737/2/89}

\bibitem[{{Dennison} {et~al.}(1998){Dennison}, {Simonetti}, \&
  {Topasna}}]{denn98a}
{Dennison}, B., {Simonetti}, J.~H., \& {Topasna}, G.~A. 1998, \pasa, 15, 147,
  \dodoi{10.1071/AS98147}

\bibitem[{{Dupuis} {et~al.}(1992){Dupuis}, {Fontaine}, {Pelletier}, \&
  {Wesemael}}]{dupu92a}
{Dupuis}, J., {Fontaine}, G., {Pelletier}, C., \& {Wesemael}, F. 1992, \apjs,
  82, 505, \dodoi{10.1086/191728}

\bibitem[{{Eggleton}(1983)}]{eggl83}
{Eggleton}, P.~P. 1983, \apj, 268, 368, \dodoi{10.1086/160960}

\bibitem[{{Falc{\'o}n-Barroso} {et~al.}(2011){Falc{\'o}n-Barroso},
  {S{\'a}nchez-Bl{\'a}zquez}, {Vazdekis}, {Ricciardelli}, {Cardiel}, {Cenarro},
  {Gorgas}, \& {Peletier}}]{falc11a}
{Falc{\'o}n-Barroso}, J., {S{\'a}nchez-Bl{\'a}zquez}, P., {Vazdekis}, A.,
  {et~al.} 2011, \aap, 532, A95, \dodoi{10.1051/0004-6361/201116842}

\bibitem[{{Fesen} {et~al.}(2015){Fesen}, {Neustadt}, {Black}, \&
  {Koeppel}}]{fese15a}
{Fesen}, R.~A., {Neustadt}, J.~M.~M., {Black}, C.~S., \& {Koeppel}, A.~H.~D.
  2015, \apj, 812, 37, \dodoi{10.1088/0004-637X/812/1/37}

\bibitem[{{Fink} {et~al.}(2007){Fink}, {Hillebrandt}, \& {R{\"o}pke}}]{fhr07}
{Fink}, M., {Hillebrandt}, W., \& {R{\"o}pke}, F.~K. 2007, \aap, 476, 1133,
  \dodoi{10.1051/0004-6361:20078438}

\bibitem[{{Fink} {et~al.}(2010){Fink}, {R{\"o}pke}, {Hillebrandt},
  {Seitenzahl}, {Sim}, \& {Kromer}}]{fink10}
{Fink}, M., {R{\"o}pke}, F.~K., {Hillebrandt}, W., {et~al.} 2010, \aap, 514,
  A53, \dodoi{10.1051/0004-6361/200913892}

\bibitem[{{Fink} {et~al.}(2014){Fink}, {Kromer}, {Seitenzahl},
  {Ciaraldi-Schoolmann}, {R{\"o}pke}, {Sim}, {Pakmor}, {Ruiter}, \&
  {Hillebrandt}}]{fink14a}
{Fink}, M., {Kromer}, M., {Seitenzahl}, I.~R., {et~al.} 2014, \mnras, 438,
  1762, \dodoi{10.1093/mnras/stt2315}

\bibitem[{{Foley} {et~al.}(2013){Foley}, {Challis}, {Chornock},
  {Ganeshalingam}, {Li}, {Marion}, {Morrell}, {Pignata}, {Stritzinger},
  {Silverman}, {Wang}, {Anderson}, {Filippenko}, {Freedman}, {Hamuy}, {Jha},
  {Kirshner}, {McCully}, {Persson}, {Phillips}, {Reichart}, \&
  {Soderberg}}]{fole13a}
{Foley}, R.~J., {Challis}, P.~J., {Chornock}, R., {et~al.} 2013, \apj, 767, 57,
  \dodoi{10.1088/0004-637X/767/1/57}

\bibitem[{{Fuller} \& {Lai}(2011)}]{fl11}
{Fuller}, J., \& {Lai}, D. 2011, \mnras, 412, 1331,
  \dodoi{10.1111/j.1365-2966.2010.18017.x}

\bibitem[{{Gaia Collaboration} {et~al.}(2018){Gaia Collaboration}, {Brown},
  {Vallenari}, {Prusti}, {de Bruijne}, {Babusiaux}, \&
  {Bailer-Jones}}]{gaia18a}
{Gaia Collaboration}, {Brown}, A.~G.~A., {Vallenari}, A., {et~al.} 2018, \aap,
  accepted (arXiv:1804.09365).
\newblock \doarXiv{1804.09365}

\bibitem[{{Gaia Collaboration} {et~al.}(2016){Gaia Collaboration}, {Prusti},
  {de Bruijne}, {Brown}, {Vallenari}, {Babusiaux}, {Bailer-Jones}, {Bastian},
  {Biermann}, {Evans}, \& et~al.}]{gaia16a}
{Gaia Collaboration}, {Prusti}, T., {de Bruijne}, J.~H.~J., {et~al.} 2016,
  \aap, 595, A1, \dodoi{10.1051/0004-6361/201629272}

\bibitem[{{G{\"a}nsicke} {et~al.}(2010){G{\"a}nsicke}, {Koester}, {Girven},
  {Marsh}, \& {Steeghs}}]{gaen10a}
{G{\"a}nsicke}, B.~T., {Koester}, D., {Girven}, J., {Marsh}, T.~R., \&
  {Steeghs}, D. 2010, Science, 327, 188, \dodoi{10.1126/science.1180228}

\bibitem[{{Geier} {et~al.}(2015){Geier}, {F{\"u}rst}, {Ziegerer}, {Kupfer},
  {Heber}, {Irrgang}, {Wang}, {Liu}, {Han}, {Sesar}, {Levitan}, {Kotak},
  {Magnier}, {Smith}, {Burgett}, {Chambers}, {Flewelling}, {Kaiser},
  {Wainscoat}, \& {Waters}}]{geie15a}
{Geier}, S., {F{\"u}rst}, F., {Ziegerer}, E., {et~al.} 2015, Science, 347,
  1126, \dodoi{10.1126/science.1259063}

\bibitem[{{Gentile Fusillo} {et~al.}(2015){Gentile Fusillo}, {G{\"a}nsicke}, \&
  {Greiss}}]{gent15a}
{Gentile Fusillo}, N.~P., {G{\"a}nsicke}, B.~T., \& {Greiss}, S. 2015, \mnras,
  448, 2260, \dodoi{10.1093/mnras/stv120}

\bibitem[{{Gilfanov} \& {Bogd{\'a}n}(2010)}]{gb10}
{Gilfanov}, M., \& {Bogd{\'a}n}, {\'A}. 2010, \nat, 463, 924,
  \dodoi{10.1038/nature08685}

\bibitem[{{Guerrero} {et~al.}(2004){Guerrero}, {Garc{\'{\i}}a-Berro}, \&
  {Isern}}]{ggi04}
{Guerrero}, J., {Garc{\'{\i}}a-Berro}, E., \& {Isern}, J. 2004, \aap, 413, 257,
  \dodoi{10.1051/0004-6361:20031504}

\bibitem[{{Guillochon} {et~al.}(2010){Guillochon}, {Dan}, {Ramirez-Ruiz}, \&
  {Rosswog}}]{guil10}
{Guillochon}, J., {Dan}, M., {Ramirez-Ruiz}, E., \& {Rosswog}, S. 2010, \apjl,
  709, L64, \dodoi{10.1088/2041-8205/709/1/L64}

\bibitem[{{Guillochon} {et~al.}(2017){Guillochon}, {Parrent}, {Kelley}, \&
  {Margutti}}]{guil17a}
{Guillochon}, J., {Parrent}, J., {Kelley}, L.~Z., \& {Margutti}, R. 2017, \apj,
  835, 64, \dodoi{10.3847/1538-4357/835/1/64}

\bibitem[{{Hansen}(2003)}]{hans03a}
{Hansen}, B.~M.~S. 2003, \apj, 582, 915, \dodoi{10.1086/344782}

\bibitem[{{Hayato} {et~al.}(2010){Hayato}, {Yamaguchi}, {Tamagawa}, {Katsuda},
  {Hwang}, {Hughes}, {Ozawa}, {Bamba}, {Kinugasa}, {Terada}, {Furuzawa},
  {Kunieda}, \& {Makishima}}]{haya10a}
{Hayato}, A., {Yamaguchi}, H., {Tamagawa}, T., {et~al.} 2010, \apj, 725, 894,
  \dodoi{10.1088/0004-637X/725/1/894}

\bibitem[{{Helder} {et~al.}(2013){Helder}, {Vink}, {Bamba}, {Bleeker},
  {Burrows}, {Ghavamian}, \& {Yamazaki}}]{held13a}
{Helder}, E.~A., {Vink}, J., {Bamba}, A., {et~al.} 2013, \mnras, 435, 910,
  \dodoi{10.1093/mnras/stt993}

\bibitem[{{Hills}(1988)}]{hill88a}
{Hills}, J.~G. 1988, \nat, 331, 687, \dodoi{10.1038/331687a0}

\bibitem[{{Hirsch} {et~al.}(2005){Hirsch}, {Heber}, {O'Toole}, \&
  {Bresolin}}]{hirs05a}
{Hirsch}, H.~A., {Heber}, U., {O'Toole}, S.~J., \& {Bresolin}, F. 2005, \aap,
  444, L61, \dodoi{10.1051/0004-6361:200500212}

\bibitem[{{Holberg} \& {Bergeron}(2006)}]{holb06a}
{Holberg}, J.~B., \& {Bergeron}, P. 2006, \aj, 132, 1221,
  \dodoi{10.1086/505938}

\bibitem[{{Iben} \& {Tutukov}(1984)}]{it84}
{Iben}, Jr., I., \& {Tutukov}, A.~V. 1984, \apjs, 54, 335,
  \dodoi{10.1086/190932}

\bibitem[{{Iben} {et~al.}(1998){Iben}, {Tutukov}, \& {Fedorova}}]{iben98a}
{Iben}, Jr., I., {Tutukov}, A.~V., \& {Fedorova}, A.~V. 1998, \apj, 503, 344,
  \dodoi{10.1086/305972}

\bibitem[{{Jordan} {et~al.}(2012){Jordan}, {Perets}, {Fisher}, \& {van
  Rossum}}]{jord12a}
{Jordan}, IV, G.~C., {Perets}, H.~B., {Fisher}, R.~T., \& {van Rossum}, D.~R.
  2012, \apjl, 761, L23, \dodoi{10.1088/2041-8205/761/2/L23}

\bibitem[{{Justham} {et~al.}(2009){Justham}, {Wolf}, {Podsiadlowski}, \&
  {Han}}]{just09a}
{Justham}, S., {Wolf}, C., {Podsiadlowski}, P., \& {Han}, Z. 2009, \aap, 493,
  1081, \dodoi{10.1051/0004-6361:200810106}

\bibitem[{{Kaplan} {et~al.}(2012){Kaplan}, {Bildsten}, \& {Steinfadt}}]{kbs12}
{Kaplan}, D.~L., {Bildsten}, L., \& {Steinfadt}, J.~D.~R. 2012, \apj, 758, 64,
  \dodoi{10.1088/0004-637X/758/1/64}

\bibitem[{{Kasen}(2010)}]{kase10}
{Kasen}, D. 2010, \apj, 708, 1025, \dodoi{10.1088/0004-637X/708/2/1025}

\bibitem[{{Kepler} {et~al.}(2016){Kepler}, {Koester}, \& {Ourique}}]{kepl16a}
{Kepler}, S.~O., {Koester}, D., \& {Ourique}, G. 2016, Science, 352, 67,
  \dodoi{10.1126/science.aad6705}

\bibitem[{{Kerzendorf} {et~al.}(2014){Kerzendorf}, {Childress},
  {Scharw{\"a}chter}, {Do}, \& {Schmidt}}]{kerz14c}
{Kerzendorf}, W.~E., {Childress}, M., {Scharw{\"a}chter}, J., {Do}, T., \&
  {Schmidt}, B.~P. 2014, \apj, 782, 27, \dodoi{10.1088/0004-637X/782/1/27}

\bibitem[{{Kerzendorf} {et~al.}(2018){Kerzendorf}, {Strampelli}, {Shen},
  {Schwab}, {Pakmor}, {Do}, {Buchner}, \& {Rest}}]{kerz18a}
{Kerzendorf}, W.~E., {Strampelli}, G., {Shen}, K.~J., {et~al.} 2018, \mnras,
  479, 192, \dodoi{10.1093/mnras/sty1357}

\bibitem[{{Kleinman} {et~al.}(2013){Kleinman}, {Kepler}, {Koester}, {Pelisoli},
  {Pe{\c c}anha}, {Nitta}, {Costa}, {Krzesinski}, {Dufour}, {Lachapelle},
  {Bergeron}, {Yip}, {Harris}, {Eisenstein}, {Althaus}, \&
  {C{\'o}rsico}}]{klei13a}
{Kleinman}, S.~J., {Kepler}, S.~O., {Koester}, D., {et~al.} 2013, \apjs, 204,
  5, \dodoi{10.1088/0067-0049/204/1/5}

\bibitem[{{Knigge} {et~al.}(2011){Knigge}, {Baraffe}, \& {Patterson}}]{knig11a}
{Knigge}, C., {Baraffe}, I., \& {Patterson}, J. 2011, \apjs, 194, 28,
  \dodoi{10.1088/0067-0049/194/2/28}

\bibitem[{{Kowalski} \& {Saumon}(2006)}]{kowa06a}
{Kowalski}, P.~M., \& {Saumon}, D. 2006, \apjl, 651, L137,
  \dodoi{10.1086/509723}

\bibitem[{{Kromer} {et~al.}(2013){Kromer}, {Fink}, {Stanishev}, {Taubenberger},
  {Ciaraldi-Schoolman}, {Pakmor}, {R{\"o}pke}, {Ruiter}, {Seitenzahl}, {Sim},
  {Blanc}, {Elias-Rosa}, \& {Hillebrandt}}]{krom13a}
{Kromer}, M., {Fink}, M., {Stanishev}, V., {et~al.} 2013, \mnras, 429, 2287,
  \dodoi{10.1093/mnras/sts498}

\bibitem[{{Kurtz} \& {Mink}(1998)}]{kurt98a}
{Kurtz}, M.~J., \& {Mink}, D.~J. 1998, \pasp, 110, 934, \dodoi{10.1086/316207}

\bibitem[{{Leonard}(2007)}]{leon07}
{Leonard}, D.~C. 2007, \apj, 670, 1275, \dodoi{10.1086/522367}

\bibitem[{{Li} {et~al.}(2011{\natexlab{a}}){Li}, {Chornock}, {Leaman},
  {Filippenko}, {Poznanski}, {Wang}, {Ganeshalingam}, \& {Mannucci}}]{li11c}
{Li}, W., {Chornock}, R., {Leaman}, J., {et~al.} 2011{\natexlab{a}}, \mnras,
  412, 1473, \dodoi{10.1111/j.1365-2966.2011.18162.x}

\bibitem[{{Li} {et~al.}(2011{\natexlab{b}}){Li}, {Bloom}, {Podsiadlowski},
  {Miller}, {Cenko}, {Jha}, {Sullivan}, {Howell}, {Nugent}, {Butler}, {Ofek},
  {Kasliwal}, {Richards}, {Stockton}, {Shih}, {Bildsten}, {Shara}, {Bibby},
  {Filippenko}, {Ganeshalingam}, {Silverman}, {Kulkarni}, {Law}, {Poznanski},
  {Quimby}, {McCully}, {Patel}, {Maguire}, \& {Shen}}]{li11}
{Li}, W., {Bloom}, J.~S., {Podsiadlowski}, P., {et~al.} 2011{\natexlab{b}},
  \nat, 480, 348, \dodoi{10.1038/nature10646}

\bibitem[{{Liu} {et~al.}(2012){Liu}, {Pakmor}, {R{\"o}pke}, {Edelmann}, {Wang},
  {Kromer}, {Hillebrandt}, \& {Han}}]{liu12a}
{Liu}, Z.~W., {Pakmor}, R., {R{\"o}pke}, F.~K., {et~al.} 2012, \aap, 548, A2,
  \dodoi{10.1051/0004-6361/201219357}

\bibitem[{{Livne}(1990)}]{livn90}
{Livne}, E. 1990, \apjl, 354, L53, \dodoi{10.1086/185721}

\bibitem[{{Long} {et~al.}(2014){Long}, {Jordan}, {van Rossum}, {Diemer},
  {Graziani}, {Kessler}, {Meyer}, {Rich}, \& {Lamb}}]{long14a}
{Long}, M., {Jordan}, IV, G.~C., {van Rossum}, D.~R., {et~al.} 2014, \apj, 789,
  103, \dodoi{10.1088/0004-637X/789/2/103}

\bibitem[{{Luri} {et~al.}(2018){Luri}, {Brown}, {Sarro}, {Arenou},
  {Bailer-Jones}, {Castro-Ginard}, {de Bruijne}, {Prusti}, {Babusiaux}, \&
  {Delgado}}]{luri18a}
{Luri}, X., {Brown}, A.~G.~A., {Sarro}, L.~M., {et~al.} 2018, \aap, accepted
  (arXiv:1804.09376).
\newblock \doarXiv{1804.09376}

\bibitem[{{Maguire} {et~al.}(2016){Maguire}, {Taubenberger}, {Sullivan}, \&
  {Mazzali}}]{magu16a}
{Maguire}, K., {Taubenberger}, S., {Sullivan}, M., \& {Mazzali}, P.~A. 2016,
  \mnras, 457, 3254, \dodoi{10.1093/mnras/stv2991}

\bibitem[{{Maoz} {et~al.}(2014){Maoz}, {Mannucci}, \& {Nelemans}}]{maoz14a}
{Maoz}, D., {Mannucci}, F., \& {Nelemans}, G. 2014, \araa, 52, 107,
  \dodoi{10.1146/annurev-astro-082812-141031}

\bibitem[{{Marietta} {et~al.}(2000){Marietta}, {Burrows}, \& {Fryxell}}]{mbf00}
{Marietta}, E., {Burrows}, A., \& {Fryxell}, B. 2000, \apjs, 128, 615,
  \dodoi{10.1086/313392}

\bibitem[{{Marsh} {et~al.}(2004){Marsh}, {Nelemans}, \& {Steeghs}}]{mns04}
{Marsh}, T.~R., {Nelemans}, G., \& {Steeghs}, D. 2004, \mnras, 350, 113,
  \dodoi{10.1111/j.1365-2966.2004.07564.x}

\bibitem[{{Nomoto}(1982{\natexlab{a}})}]{nomo82a}
{Nomoto}, K. 1982{\natexlab{a}}, \apj, 253, 798, \dodoi{10.1086/159682}

\bibitem[{{Nomoto}(1982{\natexlab{b}})}]{nomo82b}
---. 1982{\natexlab{b}}, \apj, 257, 780, \dodoi{10.1086/160031}

\bibitem[{{Olling} {et~al.}(2015){Olling}, {Mushotzky}, {Shaya}, {Rest},
  {Garnavich}, {Tucker}, {Kasen}, {Margheim}, \& {Filippenko}}]{olli15a}
{Olling}, R.~P., {Mushotzky}, R., {Shaya}, E.~J., {et~al.} 2015, \nat, 521,
  332, \dodoi{10.1038/nature14455}

\bibitem[{{Pakmor} {et~al.}(2012){Pakmor}, {Kromer}, {Taubenberger}, {Sim},
  {R{\"o}pke}, \& {Hillebrandt}}]{pakm12b}
{Pakmor}, R., {Kromer}, M., {Taubenberger}, S., {et~al.} 2012, \apjl, 747, L10,
  \dodoi{10.1088/2041-8205/747/1/L10}

\bibitem[{{Pakmor} {et~al.}(2013){Pakmor}, {Kromer}, {Taubenberger}, \&
  {Springel}}]{pakm13a}
{Pakmor}, R., {Kromer}, M., {Taubenberger}, S., \& {Springel}, V. 2013, \apjl,
  770, L8, \dodoi{10.1088/2041-8205/770/1/L8}

\bibitem[{{Pakmor} {et~al.}(2008){Pakmor}, {R{\"o}pke}, {Weiss}, \&
  {Hillebrandt}}]{pakm08a}
{Pakmor}, R., {R{\"o}pke}, F.~K., {Weiss}, A., \& {Hillebrandt}, W. 2008, \aap,
  489, 943, \dodoi{10.1051/0004-6361:200810456}

\bibitem[{{Pan} {et~al.}(2013){Pan}, {Ricker}, \& {Taam}}]{pan13a}
{Pan}, K.-C., {Ricker}, P.~M., \& {Taam}, R.~E. 2013, \apj, 773, 49,
  \dodoi{10.1088/0004-637X/773/1/49}

\bibitem[{{Pan} {et~al.}(2014){Pan}, {Ricker}, \& {Taam}}]{pan14a}
---. 2014, \apj, 792, 71, \dodoi{10.1088/0004-637X/792/1/71}

\bibitem[{{Pankey}(1962)}]{pank62a}
{Pankey}, Jr., T. 1962, PhD thesis, Howard University

\bibitem[{{Papish} {et~al.}(2015){Papish}, {Soker}, {Garc{\'{\i}}a-Berro}, \&
  {Aznar-Sigu{\'a}n}}]{papi15a}
{Papish}, O., {Soker}, N., {Garc{\'{\i}}a-Berro}, E., \& {Aznar-Sigu{\'a}n}, G.
  2015, \mnras, 449, 942, \dodoi{10.1093/mnras/stv337}

\bibitem[{{Paquette} {et~al.}(1986){Paquette}, {Pelletier}, {Fontaine}, \&
  {Michaud}}]{paqu86b}
{Paquette}, C., {Pelletier}, C., {Fontaine}, G., \& {Michaud}, G. 1986, \apjs,
  61, 197, \dodoi{10.1086/191112}

\bibitem[{{Paxton} {et~al.}(2011){Paxton}, {Bildsten}, {Dotter}, {Herwig},
  {Lesaffre}, \& {Timmes}}]{paxt11}
{Paxton}, B., {Bildsten}, L., {Dotter}, A., {et~al.} 2011, \apjs, 192, 3,
  \dodoi{10.1088/0067-0049/192/1/3}

\bibitem[{{Paxton} {et~al.}(2013){Paxton}, {Cantiello}, {Arras}, {Bildsten},
  {Brown}, {Dotter}, {Mankovich}, {Montgomery}, {Stello}, {Timmes}, \&
  {Townsend}}]{paxt13}
{Paxton}, B., {Cantiello}, M., {Arras}, P., {et~al.} 2013, \apjs, 208, 4,
  \dodoi{10.1088/0067-0049/208/1/4}

\bibitem[{{Paxton} {et~al.}(2015){Paxton}, {Marchant}, {Schwab}, {Bauer},
  {Bildsten}, {Cantiello}, {Dessart}, {Farmer}, {Hu}, {Langer}, {Townsend},
  {Townsley}, \& {Timmes}}]{paxt15a}
{Paxton}, B., {Marchant}, P., {Schwab}, J., {et~al.} 2015, \apjs, 220, 15,
  \dodoi{10.1088/0067-0049/220/1/15}

\bibitem[{{Paxton} {et~al.}(2018){Paxton}, {Schwab}, {Bauer}, {Bildsten},
  {Blinnikov}, {Duffell}, {Farmer}, {Goldberg}, {Marchant}, {Sorokina},
  {Thoul}, {Townsend}, \& {Timmes}}]{paxt18a}
{Paxton}, B., {Schwab}, J., {Bauer}, E.~B., {et~al.} 2018, \apjs, 234, 34,
  \dodoi{10.3847/1538-4365/aaa5a8}

\bibitem[{{Perlmutter} {et~al.}(1999){Perlmutter}, {Aldering}, {Goldhaber},
  {Knop}, {Nugent}, {Castro}, {Deustua}, {Fabbro}, {Goobar}, {Groom}, {Hook},
  {Kim}, {Kim}, {Lee}, {Nunes}, {Pain}, {Pennypacker}, {Quimby}, {Lidman},
  {Ellis}, {Irwin}, {McMahon}, {Ruiz-Lapuente}, {Walton}, {Schaefer}, {Boyle},
  {Filippenko}, {Matheson}, {Fruchter}, {Panagia}, {Newberg}, {Couch}, \& {The
  Supernova Cosmology Project}}]{perl99}
{Perlmutter}, S., {Aldering}, G., {Goldhaber}, G., {et~al.} 1999, \apj, 517,
  565, \dodoi{10.1086/307221}

\bibitem[{{Raddi} {et~al.}(2018{\natexlab{a}}){Raddi}, {Hollands}, {Gaensicke},
  {Townsley}, {Hermes}, {Gentile Fusillo}, \& {Koester}}]{radd18b}
{Raddi}, R., {Hollands}, M.~A., {Gaensicke}, B.~T., {et~al.}
  2018{\natexlab{a}}, \mnras, submitted (arXiv:1804.09677).
\newblock \doarXiv{1804.09677}

\bibitem[{{Raddi} {et~al.}(2018{\natexlab{b}}){Raddi}, {Hollands}, {Koester},
  {Gaensicke}, {Gentile-Fusillo}, {Hermes}, \& {Townsley}}]{radd18a}
{Raddi}, R., {Hollands}, M.~A., {Koester}, D., {et~al.} 2018{\natexlab{b}},
  \apj, accepted (arXiv:1803.07564).
\newblock \doarXiv{1803.07564}

\bibitem[{{Raskin} {et~al.}(2014){Raskin}, {Kasen}, {Moll}, {Schwab}, \&
  {Woosley}}]{rask14a}
{Raskin}, C., {Kasen}, D., {Moll}, R., {Schwab}, J., \& {Woosley}, S. 2014,
  \apj, 788, 75, \dodoi{10.1088/0004-637X/788/1/75}

\bibitem[{{Raskin} {et~al.}(2012){Raskin}, {Scannapieco}, {Fryer},
  {Rockefeller}, \& {Timmes}}]{rask12}
{Raskin}, C., {Scannapieco}, E., {Fryer}, C., {Rockefeller}, G., \& {Timmes},
  F.~X. 2012, \apj, 746, 62, \dodoi{10.1088/0004-637X/746/1/62}

\bibitem[{{Riess} {et~al.}(1998){Riess}, {Filippenko}, {Challis},
  {Clocchiatti}, {Diercks}, {Garnavich}, {Gilliland}, {Hogan}, {Jha},
  {Kirshner}, {Leibundgut}, {Phillips}, {Reiss}, {Schmidt}, {Schommer},
  {Smith}, {Spyromilio}, {Stubbs}, {Suntzeff}, \& {Tonry}}]{ries98}
{Riess}, A.~G., {Filippenko}, A.~V., {Challis}, P., {et~al.} 1998, \aj, 116,
  1009, \dodoi{10.1086/300499}

\bibitem[{{Ruiz-Lapuente} {et~al.}(2004){Ruiz-Lapuente}, {Comeron},
  {M{\'e}ndez}, {Canal}, {Smartt}, {Filippenko}, {Kurucz}, {Chornock}, {Foley},
  {Stanishev}, \& {Ibata}}]{ruiz04b}
{Ruiz-Lapuente}, P., {Comeron}, F., {M{\'e}ndez}, J., {et~al.} 2004, \nat, 431,
  1069, \dodoi{10.1038/nature03006}

\bibitem[{{S{\'a}nchez-Bl{\'a}zquez} {et~al.}(2006){S{\'a}nchez-Bl{\'a}zquez},
  {Peletier}, {Jim{\'e}nez-Vicente}, {Cardiel}, {Cenarro},
  {Falc{\'o}n-Barroso}, {Gorgas}, {Selam}, \& {Vazdekis}}]{sanc06a}
{S{\'a}nchez-Bl{\'a}zquez}, P., {Peletier}, R.~F., {Jim{\'e}nez-Vicente}, J.,
  {et~al.} 2006, \mnras, 371, 703, \dodoi{10.1111/j.1365-2966.2006.10699.x}

\bibitem[{{Sankrit} {et~al.}(2005){Sankrit}, {Blair}, {Delaney}, {Rudnick},
  {Harrus}, \& {Ennis}}]{sank05a}
{Sankrit}, R., {Blair}, W.~P., {Delaney}, T., {et~al.} 2005, Advances in Space
  Research, 35, 1027, \dodoi{10.1016/j.asr.2004.11.018}

\bibitem[{{Schaefer} \& {Pagnotta}(2012)}]{sp12}
{Schaefer}, B.~E., \& {Pagnotta}, A. 2012, \nat, 481, 164,
  \dodoi{10.1038/nature10692}

\bibitem[{{Shappee} {et~al.}(2013){Shappee}, {Kochanek}, \& {Stanek}}]{shap13a}
{Shappee}, B.~J., {Kochanek}, C.~S., \& {Stanek}, K.~Z. 2013, \apj, 765, 150,
  \dodoi{10.1088/0004-637X/765/2/150}

\bibitem[{{Shen}(2015)}]{shen15a}
{Shen}, K.~J. 2015, \apjl, 805, L6, \dodoi{10.1088/2041-8205/805/1/L6}

\bibitem[{{Shen} \& {Bildsten}(2009)}]{sb09b}
{Shen}, K.~J., \& {Bildsten}, L. 2009, \apj, 699, 1365,
  \dodoi{10.1088/0004-637X/699/2/1365}

\bibitem[{{Shen} \& {Bildsten}(2014)}]{shen14a}
---. 2014, \apj, 785, 61, \dodoi{10.1088/0004-637X/785/1/61}

\bibitem[{{Shen} {et~al.}(2013){Shen}, {Guillochon}, \& {Foley}}]{shen13a}
{Shen}, K.~J., {Guillochon}, J., \& {Foley}, R.~J. 2013, \apjl, 770, L35,
  \dodoi{10.1088/2041-8205/770/2/L35}

\bibitem[{{Shen} {et~al.}(2018){Shen}, {Kasen}, {Miles}, \&
  {Townsley}}]{shen18a}
{Shen}, K.~J., {Kasen}, D., {Miles}, B.~J., \& {Townsley}, D.~M. 2018, \apj,
  854, 52, \dodoi{10.3847/1538-4357/aaa8de}

\bibitem[{{Shen} \& {Moore}(2014)}]{shen14b}
{Shen}, K.~J., \& {Moore}, K. 2014, \apj, 797, 46,
  \dodoi{10.1088/0004-637X/797/1/46}

\bibitem[{{Shen} \& {Schwab}(2017)}]{shen17a}
{Shen}, K.~J., \& {Schwab}, J. 2017, \apj, 834, 180,
  \dodoi{10.3847/1538-4357/834/2/180}

\bibitem[{{Sim} {et~al.}(2010){Sim}, {R{\"o}pke}, {Hillebrandt}, {Kromer},
  {Pakmor}, {Fink}, {Ruiter}, \& {Seitenzahl}}]{sim10}
{Sim}, S.~A., {R{\"o}pke}, F.~K., {Hillebrandt}, W., {et~al.} 2010, \apjl, 714,
  L52, \dodoi{10.1088/2041-8205/714/1/L52}

\bibitem[{{Taam}(1980)}]{taam80b}
{Taam}, R.~E. 1980, \apj, 242, 749, \dodoi{10.1086/158509}

\bibitem[{{Taubenberger} {et~al.}(2013){Taubenberger}, {Kromer}, {Pakmor},
  {Pignata}, {Maeda}, {Hachinger}, {Leibundgut}, \& {Hillebrandt}}]{taub13a}
{Taubenberger}, S., {Kromer}, M., {Pakmor}, R., {et~al.} 2013, \apjl, 775, L43,
  \dodoi{10.1088/2041-8205/775/2/L43}

\bibitem[{{The Astropy Collaboration} {et~al.}(2018){The Astropy
  Collaboration}, {Price-Whelan}, {Sip{\H o}cz}, {G{\"u}nther}, {Lim},
  {Crawford}, {Conseil}, {Shupe}, {Craig}, {Dencheva}, {Ginsburg},
  {VanderPlas}, {Bradley}, {P{\'e}rez-Su{\'a}rez}, {de Val-Borro}, {Aldcroft},
  {Cruz}, {Robitaille}, {Tollerud}, {Ardelean}, {Babej}, {Bachetti}, {Bakanov},
  {Bamford}, {Barentsen}, {Barmby}, {Baumbach}, {Berry}, {Biscani}, {Boquien},
  {Bostroem}, {Bouma}, {Brammer}, {Bray}, {Breytenbach}, {Buddelmeijer},
  {Burke}, {Calderone}, {Cano Rodr{\'{\i}}guez}, {Cara}, {Cardoso},
  {Cheedella}, {Copin}, {Crichton}, {D{\'A}vella}, {Deil}, {Depagne},
  {Dietrich}, {Donath}, {Droettboom}, {Earl}, {Erben}, {Fabbro}, {Ferreira},
  {Finethy}, {Fox}, {Garrison}, {Gibbons}, {Goldstein}, {Gommers}, {Greco},
  {Greenfield}, {Groener}, {Grollier}, {Hagen}, {Hirst}, {Homeier}, {Horton},
  {Hosseinzadeh}, {Hu}, {Hunkeler}, {Ivezi{\'c}}, {Jain}, {Jenness}, {Kanarek},
  {Kendrew}, {Kern}, {Kerzendorf}, {Khvalko}, {King}, {Kirkby}, {Kulkarni},
  {Kumar}, {Lee}, {Lenz}, {Littlefair}, {Ma}, {Macleod}, {Mastropietro},
  {McCully}, {Montagnac}, {Morris}, {Mueller}, {Mumford}, {Muna}, {Murphy},
  {Nelson}, {Nguyen}, {Ninan}, {N{\"o}the}, {Ogaz}, {Oh}, {Parejko}, {Parley},
  {Pascual}, {Patil}, {Patil}, {Plunkett}, {Prochaska}, {Rastogi}, {Reddy
  Janga}, {Sabater}, {Sakurikar}, {Seifert}, {Sherbert}, {Sherwood-Taylor},
  {Shih}, {Sick}, {Silbiger}, {Singanamalla}, {Singer}, {Sladen}, {Sooley},
  {Sornarajah}, {Streicher}, {Teuben}, {Thomas}, {Tremblay}, {Turner},
  {Terr{\'o}n}, {van Kerkwijk}, {de la Vega}, {Watkins}, {Weaver}, {Whitmore},
  {Woillez}, \& {Zabalza}}]{astr18a}
{The Astropy Collaboration}, {Price-Whelan}, A.~M., {Sip{\H o}cz}, B.~M.,
  {et~al.} 2018, arXiv:1801.02634.
\newblock \doarXiv{1801.02634}

\bibitem[{{Timmes} {et~al.}(1995){Timmes}, {Woosley}, \& {Weaver}}]{tww95}
{Timmes}, F.~X., {Woosley}, S.~E., \& {Weaver}, T.~A. 1995, \apjs, 98, 617,
  \dodoi{10.1086/192172}

\bibitem[{{Tonry} \& {Davis}(1979)}]{tonr79a}
{Tonry}, J., \& {Davis}, M. 1979, \aj, 84, 1511, \dodoi{10.1086/112569}

\bibitem[{{Tremblay} {et~al.}(2011){Tremblay}, {Bergeron}, \&
  {Gianninas}}]{trem11a}
{Tremblay}, P.-E., {Bergeron}, P., \& {Gianninas}, A. 2011, \apj, 730, 128,
  \dodoi{10.1088/0004-637X/730/2/128}

\bibitem[{{Vennes} {et~al.}(2017){Vennes}, {Nemeth}, {Kawka}, {Thorstensen},
  {Khalack}, {Ferrario}, \& {Alper}}]{venn17a}
{Vennes}, S., {Nemeth}, P., {Kawka}, A., {et~al.} 2017, Science, 357, 680,
  \dodoi{10.1126/science.aam8378}

\bibitem[{{Wang} \& {Wheeler}(2008)}]{wang08a}
{Wang}, L., \& {Wheeler}, J.~C. 2008, \araa, 46, 433,
  \dodoi{10.1146/annurev.astro.46.060407.145139}

\bibitem[{{Webbink}(1984)}]{webb84}
{Webbink}, R.~F. 1984, \apj, 277, 355, \dodoi{10.1086/161701}

\bibitem[{{Whelan} \& {Iben}(1973)}]{wi73}
{Whelan}, J., \& {Iben}, I.~J. 1973, \apj, 186, 1007, \dodoi{10.1086/152565}

\bibitem[{{Winkler} {et~al.}(2003){Winkler}, {Gupta}, \& {Long}}]{wink03a}
{Winkler}, P.~F., {Gupta}, G., \& {Long}, K.~S. 2003, \apj, 585, 324,
  \dodoi{10.1086/345985}

\bibitem[{{Wolf} {et~al.}(2018){Wolf}, {Onken}, {Luvaul}, {Schmidt}, {Bessell},
  {Chang}, {Da Costa}, {Mackey}, {Martin-Jones}, {Murphy}, {Preston}, {Scalzo},
  {Shao}, {Smillie}, {Tisserand}, {White}, \& {Yuan}}]{wolf18a}
{Wolf}, C., {Onken}, C.~A., {Luvaul}, L.~C., {et~al.} 2018, \pasa, 35, e010,
  \dodoi{10.1017/pasa.2018.5}

\bibitem[{{Woods} {et~al.}(2017){Woods}, {Ghavamian}, {Badenes}, \&
  {Gilfanov}}]{wood17a}
{Woods}, T.~E., {Ghavamian}, P., {Badenes}, C., \& {Gilfanov}, M. 2017, Nature
  Astronomy, 1, 800, \dodoi{10.1038/s41550-017-0263-5}

\bibitem[{{Woosley} {et~al.}(1986){Woosley}, {Taam}, \& {Weaver}}]{wtw86}
{Woosley}, S.~E., {Taam}, R.~E., \& {Weaver}, T.~A. 1986, \apj, 301, 601,
  \dodoi{10.1086/163926}

\bibitem[{{Yoon} \& {Langer}(2003)}]{yoon03a}
{Yoon}, S.-C., \& {Langer}, N. 2003, \aap, 412, L53,
  \dodoi{10.1051/0004-6361:20034607}

\end{thebibliography}
\end{document}